\documentclass[review,authoryear]{elsarticle}
\usepackage{lineno,hyperref}
\usepackage{graphicx,calc,amsfonts,amssymb,sidecap}
\usepackage{setspace}
\usepackage{setspace,rotating}
\usepackage[bf]{caption}

\modulolinenumbers[5]

\journal{Physics of the Earth and Planetary Interiors}

\biboptions{authoryear}

\begin{document}

\begin{frontmatter}

\title{Detecting the oldest geodynamo and attendant shielding from the solar wind: Implications for habitability}

\author[EES,PAS]{John A. Tarduno\corref{mycorrespondingauthor}}

\author[PAS]{Eric G. Blackman}

\author[PAS]{Eric E. Mamajek}

\address[EES]{Department of Earth and Environmental Sciences, University of Rochester, Rochester, New York, 14627, USA}
\address[PAS]{Department of Physics and Astronomy, University of Rochester, Rochester, New York, 14627, USA}
\cortext[mycorrespondingauthor]{Corresponding author: Tel. 585 275 5713 fax 585 244 5689 E-mail address: john.tarduno@rochester.edu}

\begin{abstract}
The onset and nature of the earliest geomagnetic field is important
for understanding the evolution of the core, atmosphere and life on
Earth.  A record of the early geodynamo is preserved in ancient
silicate crystals containing minute magnetic inclusions. These data
indicate the presence of a geodynamo during the Paleoarchean, between
3.4 and 3.45 billion years ago. While the magnetic field sheltered
Earth's atmosphere from erosion at this time, standoff of the solar
wind was greatly reduced, and similar to that during modern extreme
solar storms. These conditions suggest that intense radiation from the
young Sun may have modified the atmosphere of the young Earth by
promoting loss of volatiles, including water. Such effects would have
been more pronounced if the field were absent or very weak prior to
3.45 billion years ago, as suggested by some models of lower mantle
evolution. The frontier is thus trying to obtain geomagnetic field
records that are $\gg$3.45 billion-years-old, as well as constraining
solar wind pressure for these times. In this review we suggest
pathways for constraining these parameters and the attendant history
of Earth's deep interior, hydrosphere and atmosphere.  In particular,
we discuss new estimates for solar wind pressure for the first 700
million years of Earth history, the competing effects of magnetic
shielding versus solar ion collection, and bounds on the detection
level of a geodynamo imposed by the presence of external fields. We
also discuss the prospects for constraining Hadean-Paleoarchean
magnetic field strength using paleointensity analyses of zircons.
\end{abstract}


\end{frontmatter}

\section{Introduction}

\vspace{-2mm} The onset and nature of the geomagnetic field is
important for understanding the evolution of the core, atmosphere and
life on Earth.  For the earliest Earth, the dynamo was powered
thermally, so dynamo startup provides information on core heat flow
and lower mantle conditions. A common posit is that the lack of
magnetic shielding of cosmic radiation is inconsistent with the
development of life, but it is clear that an atmosphere and ocean
layer can provide some protection. The primary issue explored here is
the survival of the hydrosphere. The magnetic field acts to prevent
atmospheric erosion by the solar wind.  In the case of the early
Earth, the magnetic field would have had to balance the greatly
enhanced solar wind pressure associated with the young
rapidly-rotating Sun.  The interplay between the magnetic field and
radiation from the young Sun controls the loss of light elements and,
ultimately, water and therefore may be a fundamental stage in the
development of a habitable planet.

The interaction of the solar wind and the geomagnetic field produces
Earth's magnetosphere (Figure 1), with the magnetopause ($r_{s}$)
defined as the point toward the Sun (the sub-solar point) where the
wind pressure is balanced by the magnetic field
\citep{Griessmeier2004}:\\

\begin{equation}
	r_s=\left[\frac{\mu_0f_0^2M_E^2}{4\pi^2(2\mu_0n_{sw}m_{p}v_{sw}^2+B^2_{IMF})}\right]^{1/6}
\end{equation}
\\
\noindent where $M_{E}$ is Earth's dipole moment, $n_{sw}$ is solar
wind density, $v_{sw}$ is solar wind velocity, $f_{0}$ is a
magnetospheric form factor (=1.16 for Earth), $\mu_{0}$ is the
permeability of free space, $m_{p}$ is a proton mass and $B_{IMF}$ is
the interplanetary magnetic field.

Today, Earth's magnetic field stands off the solar wind to a distance
between 10 and 11 Earth radii.  The occurrence of magnetic storms and
their associated phenomena-- including auroral lights seen at
unusually low latitudes -- are a vivid reminder that the solar input
side of this balance is variable.  In the case of a magnetic storm, a
coronal mass ejection impinges on and compresses the magnetosphere,
greatly reducing the magnetopause standoff distance.

Of course, Earth's magnetic field is dynamic and it also varies on a
wide range of timescales. At any given time, the geomagnetic field at
a radius $r$, colatitude $\theta$ and longitude $\phi$ can be
specified by the gradient of the scalar potential ($\Phi$):

\begin{equation}
\Phi(r,\theta,\phi)=r_{e}\sum_{l=1}^{\infty} \sum_{m=0}^{l}
\left(\frac{r_{e}}{r}
\right)^{l+1}   P^{m}_{l}(cos \theta)[g^{m}_{l}\cos m \phi + h^{m}_{l}\sin m
\phi]
\end{equation}

\noindent
where $P^{m}_{l}$ are partially normalized Schmidt functions, $r_{e}$
is the radius of Earth and the Gauss coefficients $g^{m}_{l}$ and
$h^{m}_{l}$ describe the size of spatially varying fields. The current
field is approximately 80\% dipolar and when averaged over several
hundred thousand years to 1 million years, the axial dipole term
($g^{1}_{0}$) is thought to dominate. The latter assumption-- the
geocentric dipole hypothesis-- forms the basis of tectonic
interpretations of paleomagnetic data that have been essential in our
understanding of the plate tectonic history of the planet.

Paleomagnetic directions provide the opportunity for tests of the
geocentric dipole hypothesis \citep{Irving1964} and these have
confirmed the hypothesis for times as old as $\sim$2.5 billion years
(Smirnov and Tarduno, 2004), notwithstanding considerable gaps in the
paleomagnetic database of directions. For even deeper time, the global
distribution of data does not yet permit a test, but we can proceed
because the geocentric dipole assumption provides the maximum magnetic
shielding \citep{Siscoe1980}.

The key magnetic parameter to determine in learning about
solar-terrestrial interactions in the deep past (eqn 1) then is
magnetic field strength ($M_{E}$), but this is far more difficult to
determine than paleodirections. Determining the ancient solar wind
pressure is also challenging, especially for the first $\sim$700
million years of Earth's history.

In this review, we will first tackle the observational challenge of
determining ancient field strength by means of examples through
Paleoarchean times. We will show how these data, when combined with
constraints based on solar analogs, allow us to constrain the
magnetopause some 3.45 billion years ago.  We provide a review of
atmospheric escape mechanisms, including contrarian views that
question the importance of a shield created by the geomagnetic
field. Within this context, we discuss the implications of the
Paleoarchean magnetopause conditions for the evolution of Earth's
water inventory.  As one peers back further in time, these data mark
the bifurcation point for potential trajectories for the dynamo. In
one scenario, the dynamo starts soon after core formation and has been
relatively strong ever since. In another, the dynamo started between 4
and $\sim$3.5 billion years ago.  We will next suggest ways to extend
the record of the magnetopause back in time to the earliest
Paleoarchean and Hadean Earth. To address the presence/absence
question about the earliest field we also address the question of
``recording zero" and in so-doing revisit the self-shielding potential
of a planet lacking a magnetic field. \\

\section{The Challenge of Recording Magnetic Deep Time}

\citet{Stevenson1983} suggested that field strength might
approximately scale with rotation, while cautioning over the accuracy
of any scaling laws of the time. Nevertheless, scaling with rotation
has been suggested as a viable way of determining past field strength
\citep{Dehant2007, Lichtenegger2010}.  While there are many unresolved
issues in dynamo theory, it is generally accepted that core convection
is driven by buoyancy \citep[e.g.][]{Roberts2000}, in which case
intensity estimates based solely on rotation are illusory. Clearly,
observations are needed.

Obtaining paleomagnetic constraints on the oldest geodynamo, however,
is difficult because of the ubiquitous metamorphism that has affected
Paleoarchean and older rocks. It is sometimes assumed that the maximum
metamorphic temperature can be simply related to the temperature at
which a magnetization is removed from a sample in the laboratory. By
this line of reasoning, if a rock had been heated to 320 $^{o}$C
during a metamorphic event, say for 1 million years, its magnetization
isolated by heating in the laboratory to 320 $^{o}$C would record the
magnetic field at the time of the metamorphic event. The magnetization
isolated as temperatures greater than 320 $^{o}$C might record the
field when the rock first cooled. In reality, many (if not most) bulk
rock samples would be totally remagnetized during this metamorphic
event, and/or their magnetic field strength records might be forever
compromised. To understand why this is so, and how we might see
through metamorphism, we must first consider the basis for measurement
of past field strength (the subdiscipline of paleointensity).\\

\subsection{Paleointensity measurement}

The physical mechanism of magnetization that is best understood is
that of thermoremanent magnetization (TRM), acquired when a sample's
constituent minerals cool through their Curie temperatures. To recover
paleointensity from rocks carrying a TRM, the most robust method is
the Thellier approach, named after seminal work of the Thelliers
\citep{Thellier1959}, or some of its more common variants
\citep[e.g. the Thellier-Coe approach;][]{Coe1967}.  In these
experiments, samples are exposed to a series of paired heating steps
(Figure 2). For example, in the Thellier-Coe approach, a sample might
first be heating to 100 $^{o}$C in field free-space, during which a
natural remanent magnetization (NRM) is removed. Next, the sample is
reheated to the same temperature, but this time in the presence of a
known applied field (this is called the ``field-on" step). After
cooling, the sample magnetization is measured and the thermoremanent
magnetization acquired can be calculated. Because one knows the
magnetization lost, the magnetization gained and the applied field
($H_{\rm lab}$), one can solve for the ancient field ($H_{\rm paleo}$)
recorded by the 100 $^{o}$C temperature range:

\begin{equation}
H_{\rm paleo} =  \frac{M_{\rm NRM}}{M_{\rm TRM}} H_{\rm lab} 
\end{equation}

\noindent
In practice, one conducts this experiment at many steps up to the maximum Curie temperature of the 
magnetic minerals in a given specimen. \\

\subsection{Chemical change}

The need for the second, or ``field-on" step makes paleointensity
experiments far less successful than a study of magnetic directions
alone. The main problem is that heating can induce changes in magnetic
minerals, changing the TRM-capacity; because the measurement is done
in the presence of a field, the reference needed for equation 3
($M_{\rm TRM}$) is compromised. In contrast, heatings for directional
measurements are always done in field-free space, and a sample can
suffer some alteration and still retain useful information.

Hence, finding samples that will not alter during laboratory
experiments is one key requirement. This constraint is also closely
related to the metamorphic and weathering history of a rock. In
general, the formation of Fe-rich clays spells disaster for
paleointensity recording because these can alter during paleointensity
experiments to form new iron-oxides \citep[e.g.][]{Cottrell2000,
  Gill2002, Kars2012}.  Ultramafic rocks, including komatiitic lavas,
might be thought of as good carriers of Archean fields, but the
Fe-rich phyllosilicate serpentine [(Mg,
  Fe)$_{3}$Si$_{2}$O$_{5}$(OH)$_{4}$] and magnetite are commonly
formed in these rocks. The magnetite carries a magnetization dating to
the alteration age (which is usually uncertain) and the magnetization
is not a TRM; instead it is a crystallization or chemical remanent
magnetization (CRM), and a straightforward tie to past field strength
is lost \citep{Yoshihara2004, Smirnov2005}.\\

\subsection{Time, Temperature and Domain State}

The principal magnetic mineral recorder of interest for Paleoarchean
and older rocks is magnetite having a Curie temperatures of $\sim$580
$^{o}$C.  Hematite, with a Curie temperature of 675 $^{o}$C is also a
potential recorder, but its occurrence as a primary phase recording a
TRM is rare relative to magnetite.  Even if we have a pristine rock
that has undergone no chemical change, magnetite within the rock can
be remagnetized by the aforementioned 320 $^{o}$C metamorphic event
even though the peak temperature is less than the Curie temperature of
magnetite. To understand this phenomena, we start with the recognition
that metamorphic heating will affect magnetic grains differently,
according to their size and related domain state.

Large magnetic grains (greater than one micron, for example), are
composed of multiple magnetic domains (called the multidomain, or MD,
state).  Even under low grade metamorphic conditions, we expect that
new magnetizations will be acquired due to movement of domain
walls. These new magnetizations will be recorded by some MD grains up
to the Curie point of magnetite.

Small magnetite grains containing only a single domain (SD) have the
potential to withstand low grade metamorphic conditions. The parameter
of interest for these grains in the thermal relaxation time ($\tau$),
which can be expressed in terms of rock magnetic parameters as follows
\citep{Dunlop1997}:

\begin{equation}
 \frac{1}{\tau}=\frac{1}{\tau_{0}} exp { \left[  - \frac{\mu_{0}V M_{s} H_{K}}{2kT} 
\left( 1- \frac{|H_{0}|}{H_{K}} \right) ^{2} \right] } 
\\[3mm]
\end{equation}
where $\tau_{0}$ (~10$^{-9}$ s) is the interval between thermal
excitations, $\mu_{0}$ is the permeability of free space, V is grain
volume, $M_{s}$ is spontaneous magnetization, $H_{K}$ is the
microscopic coercive force, k is Boltzmann's constant, T is
temperature, and $H_{0}$ is the applied field.  This relationship is
derived from N\'{e}el's \citep{Neel1949, Neel1955} theory for single
domain TRM; it was used by \cite{Pullaiah1975} to determine
time-temperature relationships that can be in turn used to predict the
acquisition of secondary magnetizations:

\begin{equation}
\frac{T_{A} ln(\tau_{A}/\tau_{0})}{M_{s}(T_{A})H_{K}(T_{A})} = 
\frac{ T_{B}ln(\tau_{B}/\tau_{0})  } {M_{s}(T_{B})H_{K}(T_{B})} 
\end{equation}

Where the two relaxation times ($\tau_{A}$, $\tau_{B}$) correspond to
temperatures ($T_{A}$, $T_{B}$) respectively, and $H_{K}>>H_{0}$. This
relationship describes the tendency for the maximum metamorphic
temperature to leak to a higher unblocking temperature range. Hence,
for our example of a peak metamorphic temperature of 320 $^{o}$C, and
a nominal reheating duration of 1 million years, only SD unblocking
temperatures up to $\sim$400 $^{o}$C should be affected
\citep{Dunlop1977}.

In summary, the challenges for paleointensity are those of finding
samples that have not been chemically altered in nature, will not
alter in laboratory, carry small single domain or single-domain like
grains (called pseudo-single PSD), and are from low metamorphic grade
geologic terrains.  The magnetic properties of common Archean rock
types vary on a grain (mineral) scale, so meeting these challenges is
rarely possible using whole rock samples.\\

\subsection{Single Crystal Paleointensity}

One approach to address these challenges is a focus on the
paleomagnetism of single crystals rather than bulk rocks, using the
single silicate crystal paleointensity (SCP) approach. Silicate
crystals are not of intrinsic magnetic interest, but they often host
minute magnetic particles that are ideal magnetic recorders
\citep{Cottrell1999, Dunlop2005, Feinberg2005, Tarduno2006}.  In
Mesozoic lavas, feldspars bearing magnetic inclusions have been shown
to alter less in the laboratory than the whole rocks from which they
were separated \citep{Cottrell2000}.  The SCP approach (using feldspar
from lavas) has been used to study geomagnetic field intensity versus
reversal frequency for the last 160 million years \citep[an inverse
  relation is supported;][]{Tarduno2001, Tarduno2002, Tarduno2005},
and during the Kiaman Superchron \citep{Cottrell2008}.  The SCP method
has also been applied to olivine from pallasite meteorites
\citep{Tarduno2012}.  The SCP approach has allowed examination of the
field of the Late, Middle and Early (Paleo) Archean as discussed
below.\\

\subsection{Prior Proterozoic-Archean Paleomagnetic Field Constraints\\}

Most prior characterizations of the Archean field have been based on
whole rocks and subsequent studies have revealed that metamorphism has
most likely compromised their paleointensity record (see discussion in
Usui et al., 2009). Below we highlight the salient observations about
the field that are robust for the interval equal to and older than the
Proterozoic-Archean boundary (Figure 3).

As noted earlier, there is a general consensus that the field was
predominately dipolar at $\sim$2.5 Ga \citep{Smirnov2004, Biggin2008,
  Smirnov2011}.  The dynamo appears to be reversing, although the
sparse rock record allows different interpretations of the reversal
rate.  \citet{Dunlop2004} and \citet{Coe2006} first argued that
reversal rates might have been low.  No long sedimentary sequences
with paleomagnetic data are available to justify this interpretation
for times before the Archean-Proterozoic boundary, and the thermal
alteration of sediments of this age questions whether such data will
ever be available. The interpretation is instead based on the sampling
of lavas that are associated with large igneous provinces exposed in
the Superior Craton of Canada \citep[the $\sim$2.5 Ga Matachewan
  Dikes;][] {Halls1991} and in the Pilbara Craton of Western Australia
\citep[the $\sim$2.7-2.8 Ga Fortesque Group;][]{Strik2003}.  However,
magmatic activity associated with these large igneous province could
have occurred in pulses (as it has during the Phanerozoic). If so,
much less time is represented by the rock record (than is expected by
assuming continuous magmatism between available radiometric age data)
and capturing only a few field reversals would be expected.

The oldest reversal documented to date is that associated with 3.1 to
3.2 Ga plutonic rocks exposed in the Barberton Greenstone Belt
\citep{Layer1998, Tarduno2007}.  An older reversal has been reported
from study of a purported 3.4 Ga tuff of the Barberton Greenstone Belt
\citep{Biggin2011}, however the sampled rocks are sandstone and minor
shale highly altered by local fluids (Axel Hofmann, personal
communication). Moreover, the limited spatial sampling and
interpretation by the authors of contamination of the data by
lightning strikes suggests this reversal record could be an artifact.

Smirnov et al. \citep{Smirnov2003} applied the SCP approach to
$\sim$2.45 Ga dikes formed close to the Proterozic-Archean boundary
from the Karelia Craton of Russia. Although the dikes were
insufficient in number to completely average secular variation, the
values are consistent with present-day field values. The SCP was also
applied to 3.2 Ga plutons of the Barberton Greenstone Belt. These
yielded a mean paleointensity that was interpreted as being within
50\% of the current field value \citep{Tarduno2007}.  However, this
relatively large uncertainty was assigned to the field value to
account for the possibility of cooling rate effects (in general,
laboratory cooling rates are by necessity much faster than those in
nature, motivating application of correction factors). Modeling and
experimental data suggest that cooling rate effects are nominal for
pseudosingle domain grains \citep{Winklhofer1997, Yu2011, Ferk2012}.
Without correction, raw data from 3.2 Ga silicate crystals are
essentially indistinguishable from the modern field intensities.

A few studies of whole rocks, where the principal magnetic carriers
were thought to be fine magnetite exsolved in silicate crystals, are
also of note. A recent restudy of the 2.78 Ga \citep{Feinberg2010}
Modipe Gabbro indicated a virtual dipole moment of 6 x 10$^{22}$ A
m$^{2}$ \citep{Muxworthy2013}.  This study lacks a field test on the
age of magnetization, but the pristine nature of select outcrops
suggests the magnetization could be primary. Inexplicably,
\citet{Muxworthy2013} do not consider the possibility of
post-emplacement tectonic tilt.  Given that the in situ magnetic
inclination is very steep (I= 70$^{o}$ reported in
\citet{Muxworthy2013}; I = 85$^{o}$ in \citet{Evans1966}, and I =
89$^{o}$ in \citet{Feinberg2010}), it is most likely that the true
paleolatitude is shallower if the Modipe Gabbro has been tilted.
\citet{Denyszyn2013} convincingly argue that the unit has been tilted,
but the exact tilting amount is uncertain.  Nevertheless, the value of
6 x 10$^{22}$ A m$^{2}$ is probably an underestimate of the true field
strength.  Study of the Stillwater Complex yielded a high
paleointensity of 92 $\mu$T \citep{Selkin2000}. After correction of
anisotropy, this value was reduced to 46 $\mu$T, and after application
of a cooling rate correction (based on SD assumptions) the value was
further lowered to 32 $\mu$T, resulting in a final VDM of 4 x
10$^{22}$ A m$^{2}$. For the reasons discussed above, the cooling rate
correction is an overestimate if PSD grains are present, suggesting
that the Stillwater VDM reported by \citet{Selkin2000} also
underestimates the true field strength.

Thus, conservatively, the hypothesis that field strength was similar
(i.e. within 50\%) to that of the modern field cannot be rejected for
the interval from the Proterozoic boundary to the mid-Archean at 3.2
Ga. Below, we discuss in more detail the constraints for field
strength for the Paleoarchean.\\

\section{The magnetopause 3.45 billion years ago}

Arguably the least metamorphosed Paleoarchean rocks are found in the
Barberton Greenstone belt of the Kaapvaal Craton, South Africa. Here,
the peak temperatures can be as low as $<$350 $^{o}$C
\citep[e.g.][]{Tice2004}.  Similar rocks are found in the closely
associated Nondeweni Greenstone Belt also of the Kaapvaal Craton,
located about 300 km south of the Barberton Greenstone Belt
\citep{Hofmann2007}.  In both belts, dacites are found which meet the
recording challenges discussed above. In a sense, they are goldilocks
recorders. They have a relatively low Fe content-- enough to record
the ancient field, but not so much that massive amounts of secondary
magnetic minerals have formed during metamorphism.\\

\subsection{Conglomerate Test}

One of the most powerful tests to determine whether a rock can retain
a primary record of the magnetic field is the conglomerate test
\citep{Graham1949} (Figure 4).  If the magnetization directions from
individual clasts forming a conglomerate are the same, that direction
must postdate deposition. If the magnetizations differ, the clasts may
retain a record of the geomagnetic field prior to deposition, and
possibly dating to the formation of the rock from which a given
conglomerate clast was derived. There is an important, and sometimes
overlooked, limitation with the application of this test.  Because
different rocks types will respond differently to a given set of
metamorphic conditions, the conglomerate test only applies to the
lithology sampled in the conglomerate test.

In the Barberton Greenstone Belt, a $\sim$3416 Ma conglomerate is
found, dominated by dacite clasts.  \citet{Usui2009} found that the
dacite clasts contained two distinct components of magnetization. At
low to intermediate unblocking temperatures, the magnetization from
many clasts defined a common direction that was indistinguishable from
the magnetic field during the $\sim$180 million-year-old Karoo
magmatic event that affected most of the Kaapvaal Craton. At high
unblocking temperatures ($\gtrsim$525-550 $^{o}$C), another distinct
component was defined; the direction of this component differed
between clasts. A formal test \citep{Watson1956} showed that these
directions indicated that the hypothesis that these directions were
drawn from a random population (as expected for a conglomerate
recording a primary direction) could not be rejected at the 95\%
confidence interval.

The parent body of the dacite clasts is exposed in the Barberton
Greenstone Belt; bulk samples show a similar component structure as
that seen in the clasts, but the high unblocking temperature
magnetization, while grouped (as would be expected in the parent rock)
showed a scatter that was unusually high as compared to that typically
seen in younger igneous rocks. It must be remembered, however, that
these are from bulk samples, which typically contain a wide range of
magnetic grain sizes. The largest multidomain grains, which are seen
in thin section \citep{Usui2009}, are expected to carry overprints and
are likely contaminating the high unblocking temperature
direction. There is also a more exotic explanation. In the absence of
a geodynamo, the high scatter could reflect a variable field produced
by solar interaction with the Paleoarchean atmosphere. For example,
the solar wind interaction with the thick atmosphere of Venus (a
planet without an internally generated magnetic field) produces an
external magnetic field which is more variable and more than an order
of magnitude weaker than the surface field of Earth \citep{Zhang2007}
(see also \textit{section 7}). Hence, paleointensity becomes the key
variable in evaluating these data further.\\

\subsection{Field Intensity Values at 3.4 to 3.45 Ga}

To obtain a field intensity at 3.45 Ga, \citet{Tarduno2010} sampled
the source of the Barberton Greenstone Belt dacite clasts; these were
supplemented with a study of 3.4 Ga dacites from the Nondweni
Greenstone Belt. In both belts, dacites contain quartz phenocrysts,
0.5 to 2 mm in size, which were the target of investigation. The ideal
nature of the magnetic inclusions within the phenocrysts was confirmed
by rock magnetic tests (e.g. magnetic hysteresis), and Thellier-Coe
paleointensity data were collected using CO$_{2}$ laser heating
methods \citep{Tarduno2007} and high resolution SQUID magnetometers
(including an instrument with a 6.3 mm access bore specialized for the
SCP approach).

The thermal demagnetization data show the removal of a low unblocking
temperature component with heating to $\sim$450 $^{o}$C (Figure
2). The unblocking characteristics of this component agree with single
domain theory suggesting that given a peak metamorphic temperature of
350 $^{o}$C, an overprint should leak to slightly higher
temperatures. At higher unblocking temperatures a component is
isolated that trends to the origin of orthogonal vector plots of the
magnetization. Over this higher unblocking temperature range,
paleointensities of 28.0 $\pm$ 4.3 $\mu$T and 18.2 $\pm$ 1.8 $\mu$T
were obtained for the Barberton and Nondweni dacites, respectively.

Because paleointensity is derived from a comparison of field-on versus
field-off steps, it does not require one to have oriented
specimens. However, oriented samples are required to obtain
paleolatitude which is in turn needed to express a given field value
at a site as a dipole moment (i.e., the value needed for eqn 1). To
obtain paleointensity, an oriented thin section approach was developed
\citep{Tarduno2007}.  Application of this technique suggests virtual
dipole moments of 3.2 $\pm$ 0.31 x 10$^{22}$ A m$^{2}$ for the
Nondweni dacite, and 4.3 $\pm$ 1.0 x 10$^{22}$ A m$^{2}$ for the
Barberton dacite. Thus a strong magnetic field was present during a
time for which we have the oldest microfossil evidence for life
\citep{Wacey2011}.  However, what were the details of the shielding
environment? Having these magnetic dipole moment constraints, the next
step in our quest to calculate magnetopause standoff is to constrain
solar wind pressure in the past, as discussed below.\\

\subsection{Solar Wind and magnetopause at 3.45 Ga}

The property of the Archean to Hadean Sun that has most captured the
attention of the geological community is its luminosity.  The Sun
began the main sequence phase of its life at a luminosity $\sim$30\%
lower than the present-day value \citep{Gough1981}.  Because of the
conversion of H to He in the Sun's core, density and temperature
increase, and luminosity rises.  The ``Faint Young Sun Paradox"
highlights the expectation that this reduced luminosity should have
resulted in a snowball Earth \citep{Sagan1972}, but sediments
(including those from the Barberton Greenstone belt) indicate
deposition in shallow warm seas.

The solution to the paradox is generally posed in the form of some
greenhouse gas concentration in the early Earth's atmosphere. We will
briefly revisit this question later, but for now the variable of
greatest interest for addressing equation 1 is not luminosity, but
other aspects of the Sun's radiation, specifically the solar wind
(mainly hydrogen ionized to protons), X-rays and energetic UV
radiation.  Unlike Earth, where field strength relates to buoyancy
forces, the Sun's magnetic field is a strong function of its rotation
rate.  The solar dynamo has its origin in the interactions between
rotation and its convective outer envelope \citep{Parker1970}.  The
Sun sheds angular momentum through the emission of magnetized winds
and rotation slows in a process known as magnetic braking. The
rotation period slows as $\sim t^{\frac{1}{2}}$
\citep[e.g.][]{Ayres1997}.

Stellar winds are not detected directly, but through interaction with
the interstellar medium.  The interaction forms an ``astrosphere"
surrounding a star created by hot, interstellar neutral hydrogen which
is heated by the stellar wind.  Hot hydrogen atoms can be detected
spectroscopically because they produce an absorption feature (H
{\small I} Lyman-$\alpha$) thereby proving an indirect means of
gauging stellar winds.

Observational constraints on ancient mass are available from the study
of solar analogs - stars that are approximately the same size as the
Sun but have different ages \citep[e.g.][]{Wood2006}.  These data
suggest that current and past mass loss are related by a power law:

\begin{equation}
	\frac{\dot{m}v_{sw}}{\dot{m}_0v_{sw0}}=\left(\frac{t}{t_0}\right)^{-2.33}
\end{equation}

\vspace{1mm}
\noindent
where $\dot{m}_{0}$, $v_{sw0}$ are the present-day mass loss and solar
wind velocity, respectively. As we will discuss later, this
relationship should be applied to the Sun only for solar ages greater
than $\sim$ 700 million years.  This relationship suggests that the
solar mass loss at 3.45 Ga was 2.4 x $10^{-13}$ M$_{\odot}$/yr, where
M$_{\odot}$ is the current solar mass.  This mass loss can be used to
further derive a model (Model A of \citet{Tarduno2010}) of wind
velocity and density change with time, which comprise solar wind
pressure.  Hence, we can now solve equation 1 to obtain estimates of
magnetopause standoff distance as a function of dipole moment (Figure
5).

The paleointensity estimates for 3.4-3.45 Ga from the Kaapvaal Craton
dacites suggest standoff distances of $\sim$5 Earth radii (R$_{e}$).
One can also estimate mass loss from rotation and X-ray emission.
Stellar evolution models predict the Sun at 3.45 Ga would appear to be
a G6V star having a rotational period of $\sim$12 days
\citep{Mamajek2008}.  Mass loss rates among solar-type stars can be
related to X-ray emission ($f_{X}$) and rotation as:
 \begin{equation}
	\dot{m}=\dot{M_{\odot}}\left(\frac{\rm{R}}{\rm{R}_{\odot}}\right)^{2}\left(\frac{f_{X}}{f_{X_{\odot}}}\right)^{1.34\pm0.18}	
\end{equation}
\\ where R is solar radius, R$_{\odot}$ is the modern value, and
$f_{X_{\odot}}$ is the modern soft X-ray surface flux
($\sim$10$^{4.57}$ erg/s/cm$^{2}$ in 0.1-2.4 keV band).  Equation 7
(Model B of \citet{Tarduno2010}) predicts a higher mass loss of 1.5 x
10$^{-12}$ M$_{\odot}$/yr and a smaller standoff of $\sim$4 R$_{e}$.
Even using the more conservative mass loss of Model A, the standoff
distance is expected to be only $\sim$50\% of the present-day value.
\citet{Sterenborg2011} presented results of a magnetohydrodynamic
numerical simulation of Palearchean conditions utilizing modules of
the Space Weather Modeling Framework \citep{Toth2005}.  They obtained
standoff distances within error of our estimates, confirming the
compressed Paleoarchean magnetosphere.

The change in standoff relative to today may in itself seem abstract,
but we have a very good idea of what these conditions are like because
they are typically experienced during modern events, such as the
Halloween solar storms of 2003 \citep[e.g.][]{Rosenqvist2005}.
However, during modern events like the Halloween storms, reduced
standoffs occur on hour to day timescales. In contrast, these
conditions would represent the typical day 3.4 to 3.45 billion years
ago.  Thus, while the magnetic field was present and provided some
shielding from the solar wind, the reduced standoff conditions suggest
there nevertheless could have been important modifications of the
early Earth's atmosphere. Below, we first provide context through a
discussion of atmospheric loss processes.\\

\section{Atmospheric Loss Mechanisms}

Atmospheric loss mechanisms are typically divided into thermal escape
stimulated by energetic photons and non-thermal escape associated with
the interaction of charged particles. Following this classification,
the presence/absence of a magnetic field will influence only
non-thermal escape. However, thermal processes are also of importance
because they determine whether non-thermal loss is likely to be
efficient.  Moreover, as we introduce below, under reduced standoff
conditions of the early Earth, heating due to solar wind interaction
with the atmosphere is possible, and in this case the clear-cut
assignment of magnetic field influence to only non-thermal processes
becomes less meaningful.

Much can be learned about these atmospheric loss processes by study of
planets without internally generated magnetic fields, namely Venus and
Mars, and the general interest has expanded in parallel with exoplanet
discoveries and considerations of habitability
\citep[e.g.][]{Seager2010}. The limitations of these analogies should
also be kept in mind; atmospheric loss processes presently working at
Venus and Mars are the result of different evolutionary histories of
their respective atmospheres. While analogies should not be taken too
far, the physical processes are illustrative, notwithstanding the
difficult task of deciding the importance of these processes for the
young Earth.\\

\subsection{Thermal escape}

The uppermost tenuous part of the atmosphere is known as the
exosphere; it begins at the exobase, where the mean free path of
molecules is so large that collisions before escape are unlikely.  In
Jeans' escape \citep{Jeans1925}, a molecule gains sufficient energy to
reach its escape velocity. The the important Jeans' escape parameter
$\lambda_{J}$ is defined as the ratio of gravitational potential
energy to thermal energy:

\begin{equation}
\lambda_{J} \equiv GMm/kT_{J}r_{J}
\end{equation}

\noindent
where, G is the gravitational constant, $M$ is the planetary mass, $m$
is the mass of the escaping particles, $k$ is Boltzmann's constant,
$T_{J}$ is the temperature of the exobase and $r_{J}$ is the radius of
the exobase.  A central parameter in understanding thermal escape in
the past is thus knowing the temperature at the exobase, and this in
turn will depend of the solar flux, particularly that in the X-ray to
energetic ultraviolet (EUV) spectral range (wavelength $\leq$ 1027
\AA) commonly referred to as XUV radiation.  Astrophysical
observations of solar analogs \citep[e.g.][]{Dorren1994} have provided
data to constrain the past solar flux.  \citet{Ribas2005} define a
flux as a function of stellar age ($t$) as:

\begin{equation}
F = 29.7\, t^{-1.23}\, {\rm ergs} \, {\rm s}^{-1} \, {\rm cm}^{-2}
\end{equation}

\noindent
where the flux is defined at wavelengths between 1 and 1200 \AA. At
2.5, 3.45 and 4 Ga, this flux would be about 3, 6, and 13 times
greater (respectively) than today. This increased flux has the
potential to heat the exosphere (but this also depends on atmospheric
composition, see {\it section 4.3}) and increase ionization, promoting
non-thermal escape.

If heating is sufficiently great ($\lambda_{J} <$1.5;
\citet{Opik1963}) atmospheric ``blowoff" can occur.  The more general
term is hydrodynamic escape, where the thermal energy from XUV
radiation allows an escaping species to efficiently stream away from
the planet. It is particularly important for hydrogen-rich exospheres,
potentially including that of the early Earth \citep{Watson1981}.
Hydrodynamic escape will be important only if the supply of the
escaping species is not limited by diffusion lower in the atmosphere.

The diffusion limited flux ($F_{d}$) can be expressed as
\citep{Hunten1973}:

\begin{equation}
F_{d}= b\left(\frac{GM}{kT}\right)(m_{j}-m_{i})\left(\frac{n_{i}}{n_{j}}\right)
\end{equation}

\noindent
where ($F_{d}$) is diffusive flux per steradian per second, $b$ is a
binary collision parameter,
\begin{math}
\cal{O} 
\end{math}
(10$^{9}$) cm$^{-1}$ s$^{-1}$, $n$ is number density and the
subscripts $i$ and $j$ refer to the escaping and background gas
masses, respectively, and $n_{i}/n_{j}$ is the mixing ratio of the
escaping gas.

If we consider hydrogen as the gas escaping from a background of
nitrogen, mixing ratios greater than a few percent suggest that escape
will be energy related rather than diffusion limited
\citep{Watson1981}, and this seems likely given the potential sources
of hydrogen ranging from dissociation of water vapor to volcanic
outgassing. In fact, faced with the likelihood of hydrogen
accumulation in a pre-Proterozoic anoxic Earth, the relevant question
is one of determining how large amounts of hydrogen were removed such
that the atmosphere did not become highly reducing (evidence for which
is lacking in Paleoarchean and younger rocks).

Two final aspects of hydrogen hydrodynamic escape are of
note. Hydrodynamic escape of hydrogen has the potential to carry out
with it heavier atmospheric species.  \citet{Watson1981} used equation
10 to predict that heavier species would probably not be removed, with
the exception of deuterium. This prediction is consistent with the
lack of nitrogen isotope ($^{15}$N/$^{14}$N) fractionation observed on
Earth \citep{Marty2013}, versus, for example, that observed on Mars
\citep[e.g.][]{Jakosky1994} where thermal and nonthermal processes
have greatly thinned the atmosphere.  However, the removal of hydrogen
could leave behind excess atmospheric oxygen that would have to be
either incorporated into the crust/mantle or lost.
\citet{Kulikov2007} suggest oxygen could be lost by hydrodynamic
escape given early Earth EUV conditions.

Another type of thermal loss is that associated with large
impacts. The Paleoarchean of the Barberton Greenstone Belt (and the
Pilbara Craton) contains evidence for large impacts in the form of
spherule beds \citep[e.g.][]{Lowe2003}; these may represent the tail
of a distribution of large impacts known as late heavy bombardment
\citep{Bottke2012}.  During a large impact the atmosphere can be
super-heated by shock to the point that some is blown-off.  We include
this process in the discussion because during this heating,
non-thermal removal processes will also be enhanced. This effect would
be most relevant for times after the main thermal pulse resulting from
shock.\\

\subsection{Non-thermal escape}

Non-thermal processes encompass a wide range of escape mechanisms that
are summarized in Table 1. Some of these involve photons
(e.g. photodissociation) whereas others involve interaction with the
charged solar wind and collectively can be grouped into ``solar wind
erosion".  Light atmospheric species that have been ionized by EUV
photons, charge exchange with solar wind protons, or by impacts with
solar wind electrons \citep{Luhmann1992} can be removed by the solar
wind. Ionized light and heavier ions that are not removed can reenter
the atmosphere. When they impact neutral species near the exobase,
some of the target species will be scattered, gaining enough energy to
escape, in a process known as ``sputtering".  Heavier species
(e.g. N$_{2}^{+}$, O$_{2}^{+}$) will be more effective as sputtering
agents \citep{Johnson1990}.

Of particular importance for our consideration are those non-thermal
processes that involve ion escape, because these generally require
open field lines. Ions can also travel along the magnetic field and
escape through the magnetotail \citep[e.g.][]{Kistler2010}.  Open flux
also occurs near the magnetic poles, where magnetic field lines
connect with the interplanetary field. The polar cap is defined as the
region where open flux occurs and the auroral zone is at its
boundary. The solar wind can penetrate deeper into the atmosphere in
the polar cap causing ionization, and this is also the region where
energetic ions can escape; for example depletion of H and He above the
Earth's magnetic poles today is due to escape along open field lines
\citep{dePater2001}. Ions can also be lost when reconnection takes
place, when the magnetic field of the solar wind joins the terrestrial
magnetic field \citep[e.g.][]{Lyon2000}. This is classically expressed
as a southward interplanetary (solar wind) magnetic field connecting
with a northward magnetosphere field, and this type of reconnection is
pronounced during coronal mass ejection events when magnetized solar
plasma impacts the magnetosphere. However, reconnection is a complex
phenomena that has been observed under a variety of solar forcing
conditions.  \\

 \newpage 
 
 \begin{table}
 \caption{ Atmospheric Escape Processes$^{\dag}$}
 \begin{tabular}{llll}
  Process & Example Reactions  & Current Planetary Example  &Early Earth$^{\ddag}$\\
 \hline \hline 
Charge exchange$^{1}$                    & H + H$^{+*}$ $\rightarrow$ H$^{+}$ + H$^{*}$  &Earth (H, D)   &$\uparrow$\\
                                                           & O + H$^{+*}$ $\rightarrow$ O$^{+}$ + H$^{*}$ &Venus (He) \\
                                                           \hline 
Dissociative recombination$^{2}$      &   O$_{2}^{+}$ + $e$ $\rightarrow$ O$^{*}$ + O$^{*}$ &  Mars (O) &$\uparrow$ \\
                                                            &     OH$^{+}$ + $e$ $\rightarrow$ O + H$^{*}$  & Venus (H), Mars (N)    \\            \hline                                          
Impact dissociation$^{3}$           &  H$_{2}$ + $e^{*}$ $\rightarrow$ H$^{*}$ + H$^{*}$     & Mars (N)   &$\uparrow$  \\                   
                                                    &  N$_{2}$ + $e^{*}$ $\rightarrow$ N$^{*}$ + N$^{*}$     &  \\ \hline
 Photodissociation                            &   O$_{2}$ + $h \nu$ $\rightarrow$ O$^{*}$ + O$^{*}$      &    &$\uparrow$              \\
                               &   H$_{2}$O + $h \nu$ $\rightarrow$ OH$^{+}$ + H$^{*}$   &   &                \\
                              &   OH$^{+}$ + $h \nu$ $\rightarrow$ O + H$^{*}$  &     &                \\ \hline
Ion-neutral reaction$^{4}$                     &   O$^{+}$H$_{2}$ $\rightarrow$ OH$^{+}$ + H$^{*}$     &    &  $\uparrow$   \\        \hline
Sputtering or               &                                                                                       &        &  $\uparrow$        \\
 knock-on$^{5}$                            &     O$^{*}$ + H $\rightarrow$ O$^{*}$ + H$^{*}$    &  Venus (H) &         \\ \hline
Solar-wind pickup$^{6}$                      &      O +   $h \nu$ $\rightarrow$    O$^{+}$ + $e$                &    Mercury (He, Ar)         & $\uparrow^{\S}$       \\
                                                            & then O$^{+}$ picked up\\ \hline
 Ion escape$^{7}$              & H$^{+*}$ escapes     &Earth (H,D, He)        & $\uparrow^{\S}$\\ \hline 
Electric fields$^{8}$        & X$^{+}$ + eV $\rightarrow$ X$^{+*}$       &Earth (H,He) & $\uparrow^{\S}$\\ 
 \hline \hline
Jeans escape          &                                                                           &                        & $\uparrow$\\
Hydrodynamic escape &                                              &                        & $\uparrow$\\
 Impact erosion          &                                                                           &                        & $\uparrow$\\
 \hline \hline
\end{tabular}
\vspace{-7mm}
\end{table}

\noindent
{\small $^{\dag}$Adapted from \citep{Hunten1989, dePater2001,
    Seager2010}.

\noindent
$^{*}$ Excess kinetic energy.

\noindent
$^{\ddag}$Referenced at $\sim$3.45 Ga for comparison with magnetopause
reported in \citep{Tarduno2010}.

\noindent
$\uparrow$, $\downarrow$, escape greater or lower at $\sim$3.45 Ga
than today.

\noindent
$\uparrow^{\S}$, overall escape greater due to increased ion supply.

\noindent
$^{1}$Charge exchange occurs when an energetic ion collides with a
neutral, and the ion loses its charge but retains its kinetic
energy. The former ion can have enough kinetic energy to escape.

\noindent 
$^{2}$Dissociative recombination occurs when an ion dissociates on
recombination with an electron, resulting in atoms with sufficient
energy to escape.

\noindent
$^{3}$Impact dissociation occurs when a neutral molecule is impacted
by an electron; the resulting energetic atoms can escape. Similar end
products result from photodissociation, but rather than an electron
impacted by a molecule, energy is transferred by photons.

\noindent
$^{4}$Ion and neutral molecule can also react to form an ion and an
energetic atom.

\noindent
$^{5}$When an energetic atom of ion collides with an atmospheric atom,
the atom is accelerated and may escape.  A single collision is
referred to as knock-on. Sputtering describes a cascade of collisions
following the initial collision, with these later collisions supplying
enough energy for atmospheric escape.

\noindent
$^{6}$In the absence of an internally generated dynamo field (or in
the case of a very weak field), charged atmosphere particles can
interact directly with the solar wind and be carried away, in a
process know as solar wind pickup or sweeping. If a planet lacks an
internal field, but has an atmospheric and an induced magnetic field,
particles are picked up at the subsolar point and lost at the tails of
the induced magnetosphere.

\noindent
$^{7}$When magnetic field lines are open, energetic atmospheric ions
can escape (ion escape).

\noindent
$^{8}$Charged particles can be accelerated by electric fields of
magnetospheric or ionospheric origin aiding escape as they are lost as
they move out along magnetic field lines (associated with either an
internal-generated or induced magnetosphere). }

\subsection{Atmospheric chemistry}

The type of loss that is most important at a given time is a function
of atmospheric chemistry. Below we focus on a few aspects of this
complex topic relevant to our considerations of atmospheric lost on
the early Earth and refer readers to the excellent synthesis on
planetary atmospheres by \citet{Pierrehumbert2010}.

The extent of the atmospheric and loss from its top will be dependent
on temperature, which in turn depends on whether the dominant species
are good or poor infrared emitters \citep{Lammer2013}.  For example,
CO$_{2}$ is a good infrared emitter; it dominates the exobase of Venus
and as a result the temperature is relatively cool (200-300 K). In
contrast, atomic O is a poor infrared emitter; it dominates the
present exobase of Earth and the temperature is relatively high (1000
K). Nitrogen is neither a good or poor IR-emiter, so the exobase
temperature will tend to be controlled by other species
\citep{Pierrehumbert2010}.  The question of what these species might
be in the early Earth's atmosphere is itself hotly debated,
principally focused on the greenhouse gasses needed to counter the
Faint Young Sun, and the amount to which the atmosphere was reducing
\citep{Shaw2008}.

Proposed solutions to the Faint Young Sun have included increased
atmospheric NH$_{3}$ \citep{Sagan1972}, CO$_{2}$ \citep{Owen1979},
CH$_{4}$ \citep{Kiehl1987, Catling2001}, or a decreased albedo
\citep{Rosing2010}.  Various counterarguments have been made (see
comprehensive review by \citet{Feulner2012}).  NH$_{3}$ should be
quickly dissociated by photolysis; its survival was postulated because
of the sheltering effect of a hypothetical thick Titan-like organic
haze layer resulting from photolysis of CH$_{4}$.  However, a haze
layer can have a cooling effect \citep{Haqq2008}. Nevertheless, an
optically thin haze layer might be an effective UV screen
\citep{Hasenkopf2011}.  The ratio of the mixed valence state magnetite
to siderite in banded iron formations (BIFs) appears to limit
atmospheric CO$_{2}$ levels \citep{Rosing2010}, but this strictly
applied to the oldest well-preserved Archean BIFs.  Moreover, this
apparent limitation relies on the assumption that iron mineral
formation in BIFS was not far from equilibrium with the atmosphere;
this assumption has been questioned \citep{Dauphas2011, Reinhard2011}.
\citet{Goldblatt2011} argue that albedo effects alone cannot resolve
the Faint Young Sun Paradox to a factor of 2 of the needed radiative
forcing (but see also responses by \citet{Rosing2011}).  Nitrogen
itself, while not a greenhouse has, could have aided early greenhouse
warming because its presence broadens the absorption lines of other
greenhouse gases \citep{Goldblatt2009}.

Notwithstanding the continuing debates, a N$_{2}$-CO$_{2}$ atmosphere,
with important time dependent roles for hydrogen and methane, seems to
be the closest consensus available for the early Earth.  The
importance of methane likely tracks the start of its production
biologically \citep{Catling2001}, whereas the very early presence of
hydrogen in the atmosphere seems unavoidable given early high
degassing rates, although concentrations are also debated
\citep{Tian2005, Catling2006}.  Several authors have noted that the
removal of hydrogen by thermal and non-thermal processes could have
contributed to the change from a reducing to oxidizing atmosphere
\citep{Catling2001, Lammer2008}.  We recall that loss of this hydrogen
can adiabatically cool the exosphere, limiting loss of heavier species
\citep{Tian2008, Lammer2013} whereas H$_{2}$-N$_{2}$ collision-induced
warming is another mechanism that could counter effects of the Faint
Young Sun \citep{Wordsworth2013}.

There is another potential bottleneck of importance that could control
the loss of water from the early Earth. Today, water loss is limited
by the cold trap at the top of the troposphere where temperatures are
low and water condenses, leaving the stratosphere relatively
dry. Given a current loss of H from the atmosphere of 2.7 $\times$
10$^{8}$ atoms cm$^{-2}$ s$^{-1}$, only about 5.7 m of a global ocean
on Earth would be removed in 4.5 billion years \citep{Hunten1989}.

Since water vapor is controlled by surface temperature, and Archean
surface temperatures are argued to have been temperate
\citep[e.g.][]{Blake2010}, there is no reason to suspect the lower
atmosphere was unusually dry.  However, with the lack of a ozone layer
in the anoxic Earth, greater penetration of UV into the atmosphere
(causing photolysis of water), and greater UV output of the young Sun,
a cold trap may not be as limiting, as has been suggested for the
primitive Venusian atmosphere \citep{Chassefiere1997}.  The example of
Venus merits further discussion (below) because it and Mars are
sometimes offered as examples relevant to the importance of magnetic
fields on atmospheric loss.\\

\subsection{The apparent Martian and Venusian counter-examples: issues in assessing the effectiveness of a magnetosphere}

While it has been traditionally postulated that the difference between
the current water inventory of Earth versus that of Mars is due to the
presence of an internally generated magnetic field on the former
\citep[e.g.][]{Lundin2007}, there have been several more recent
dissenting voices, to the point that some have even questioned the
more general importance of a magnetic field \citep{Kasting2010} and
particularly whether a stronger field could even lead to increased
dispersal of the atmosphere \citep{Brain2013}.  These issues raised,
follow one or more of the following threads of reasoning:\\

\noindent
{\it (i)} Venus lacks an internal magnetic field but retains a thick
atmosphere, providing a circumstance in which seemingly magnetic
shielding is unimportant for atmospheric protection, and by extension,
habitability. \\

\noindent
{\it (ii)} The present-day total atmospheric ion escape from Earth,
Mars and Venus appears to be within similar orders of magnitude, at
10$^{24}$-10$^{26}$ ions s$^{-1}$ \citep{Barabash2010,
  Strangeway2010}. Because Earth has an internal geomagnetic field and
Mars and Venus do not, an internal magnetic field must not be that
important for water loss/survival.\\

\noindent
{\it (iii)} Planets with atmospheres have an induced magnetosphere
that provides atmospheric protection, therefore a specifically
internally generated magnetic field is unnecessary for atmospheric
survival.\\

\noindent
{\it (iv)} Ions can be accelerated and channeled along large scale
field lines, particular toward the poles in a large scale dipole
configuration. In this respect, by converging the incoming particle
flux, the magnetic fields help to exacerbate the atmospheric loss
rather than abate it \citep{Brain2013}\\

Sorting out the evidence in support of and in opposition to these
arguments comprises an important opportunity for further research.
Here we discuss each of them a bit further to highlight what we think
are particularly important aspects to consider further when critically
assessing the role of magnetic fields in atmospheric retention and
habitability.

With regard to argument {\it i}, the distinction between merely
retaining an atmosphere vs. the retention of specifically water must
be kept in mind. While the exact evolutionary history and escape
mechanisms are debatable \citep[e.g.][] {Kasting1988,
  Elkins-Tanton2013, Hamano2013}, there is little argument that Venus
has lost a huge amount of water relative to Earth. This has
irreversibly altered its atmosphere, to the point that the probative
value of present-day conditions on Venus for understanding the
solar-terrestrial environment of the early Earth is limited. The
present Venusian CO$_{2}$ atmosphere has a relatively cool exobase
limiting loss. It is true that a magnetic field is not necessary to
retain an atmosphere but, the relevant issue for life as we know it is
water.

The evolution of a a planetary atmosphere is also dependent on the
evolution of its host star's wind.  Indeed the fact that the young sun
had a stronger solar wind is important in considering argument {\it
  ii}, which ignores the greatly enhanced solar winds and XUV
radiation environment associated with the rapidly rotating young Sun
({\it sections 3.3, 4.1} of this paper).  In fact the extreme XUV
atmospheric heating may have been so great that the exosphere moved
beyond the magnetosphere associated with a geodynamo
\citep[e.g.][]{Lammer2008, Tian2008}. In this case, by definition part
of the atmosphere will be unprotected by the magnetic field against
loss. But atmospheric escape rates could be even higher in such a case
in the absence of an internally generated magnetic field, because
solar wind protons would more readily gain access to denser parts of
the atmosphere. In fact, the question becomes one of atmospheric
survival, even given a N$_{2}$-CO$_{2}$ atmosphere
\citep{Lichtenegger2010}.  Recent measurements of present-day
conditions do show that Mars and Venus have in fact exhibited
accelerated atmospheric loss during times of increased solar wind
pressure \citep{Edberg2010, Edberg2011}, whereas a direct comparison
of Earth and Mars under the same solar forcing found that Martian
atmospheric loss was more sensitive to increases in solar wind
pressure \citep{Wei2012}, strongly suggesting the importance of the
geomagnetic shield.

In assessing the importance of a magnetic field for habitable
atmosphere retention, we must in fact distinguish between a field
induced externally by wind-atmosphere interactions (argument {\it
  iii}) vs. supplied internally via a dynamo.  The structure and
location of influence of the field would be different in the two cases
and this must be considered in comparing the relative influence.  As
summarized by \cite{Brain2013}, the field arising from interactions
between the wind (perhaps as in present day Venus, Mars and Titan)
likely deflects the wind only within $\lesssim 1$ planetary radii
where as in internally produced field could be stronger, and abate the
incoming wind are larger radii from the planet.

Because an internally produced magnetic field would be anchored in the
core, field lines converge toward magnetic poles.  The role of a large
magnetosphere combined with convergence of the field toward the poles
may indeed be important in assessing the sign of the influence of the
magnetic field on atmosphere retention, as suggested in argument {\it
  iv}.  Because incoming wind ions typically have very small
gyro-radii, they are captured onto field lines that are concentrated
toward the poles as discussed in \cite{Brain2013}. Supporting evidence
comes in part from observations \cite{Strangeway2005} of the outward
flux of O ions from auroral zones is consistent with concentration of
a higher solar wind energy flux by up to 2 orders of magnitude in
these regions, and is consistent with theoretical considerations of
\cite{Moore2010}.  Such an effect would preferentially eject heavy
ions that would otherwise be gravitational bound at atmospheric
temperatures, unlike light ions like H$^{+}$ whose thermal speed would
already exceed the escape speed.

Despite the plausibility of this magnetically aided concentration of
wind flux toward the polar caps, we note that there are a number of
effects to keep in mind in assessing its influence on habitable
atmosphere retention.  First, note that the same magnetic fields that
might help concentrate the effect of the solar wind in local regions,
also create a magnetospheric environment that can result in
recapture. For example, the polar outflow of O$^{+}$ from Earth today
is some 9 times greater than the net loss, considering recapture in
the magnetosphere \citep{Seki2001}. A local outflow might therefore
not always result in a catastrophic global outflow.  Second, the depth
to which ions can be accelerated by conversion of solar wind flux into
the accelerating Poynting flux also depends on competing forces such
as a magnetic mirror force \citep{Cowley1990}.  Converging field lines
can reflect particles, and trap them above a certain height.  The
specific depth to which ions can penetrate may be limited for specific
planetary circumstances and should be quantified. Third, the time
scale for the full atmosphere of ions essential for habitability to be
ejected at local auroral regions depends on how fast the atmosphere
can circulate into these regions of loss.  This resupply requires
horizontal flow to refill the cone of loss and thermal diffusion to
move material up to the exobase. The slower of these mechanisms
determines the rate of loss.

While our main purpose here is to provide an introduction to the
issues warranting further quantitative study, we now present one
calculation on which all influences of the magnetic fields depend,
namely the estimate of the impinging rate of solar ions for
magnetically shielded versus unshielded planets (Figure 6).
Regardless of how the flux that penetrates the atmosphere subsequently
evolves, this calculation is the basic starting point for assessing
how much stellar wind flux first enters the magnetosphere.

The rate of solar ions with the potential to enter a planet's
atmosphere depends on the product of the effective area onto which
ions are captured and the speed of the inflow. A large magnetosphere
increases the effective area but reduces the inflow speed at which the
ions flow into the atmosphere by abating the ram pressure of the solar
wind and compressing the solar wind field at the magnetopause with the
solar wind ions kept out.  However, when the solar wind and
magnetosphere fields reconnect, the solar wind ions can bleed onto the
planetary magnetic field and into the atmosphere.  The speed at which
ions flow into the atmosphere is $\sim$1/2 the inflow reconnection
speed when averaged over time if we assume that the solar wind field
orientation is favorable to reconnection $\sim 50\% $ of the
time. During the reconnection phase, the effective collecting area of
stellar wind ions can become as large as the magnetopause radius (on
the daylit side).  Keeping these concepts in mind, the mass flux
collected by a planet without a magnetosphere ${\dot M}_{c}$ from a
stellar wind whose velocity exceeds the escape speed of the planet at
its surface can be written as:
\begin{equation}
{\dot M}_{c}\simeq \frac{\pi r_p^2}{4\pi R^2}{{\dot M}_w}
\label{4.4.1}
\end{equation}
where $r_p$ is the planet radius and $R$ is the planet star distance
assuming $R>>r_p$, and ${\dot M}_w$ is the wind mass flux. In the
presence of a magnetosphere, the mass flux collected ${\dot M}_{c,m}$
is given by:
\begin{equation}
{\dot M}_{c,m}\simeq \frac{\pi r_{m}^2 }{4\pi R^2 } \frac{v_{rec}/2} {v_{sw}}{{\dot M}_w}  
\label{4.4.2}
\end{equation}
where $v_{rec}$ is the reconnection speed at the magnetopause,
$v_{sw}$ is the wind speed, and $r_s$ is the magnetospheric distance
from the planet. We have assumed $R>>r_s$.  Taking the ratio of
equation 12 to equation 11 gives:
\begin{equation}
Q\equiv {{\dot M}_{c,m}\over {\dot M}_{c}}\simeq 1.75 \left (v_{rec} \over 25 {\rm km/s}\right )  \left (v_{sw} \over 400 {\rm km/s}\right )^{-1}  \left ({r_s/r_p \over 10}\right)^2  
\label{4.4.1}
\end{equation}
where we have scaled to the present solar wind- Earth values for
reconnection at the magnetopause
\citep{Cassak2007,Paschmann2013,Walsh2013}. Note that the estimate of
the reconnection speed at the magnetopause is consistent both with the
value estimated theoretically based on asymmetric reconnection
\citep{Cassak2007} and observations \citep{Walsh2013}.

While a crude estimate, if we scale equation 13 to Earth at 3.45 Ga
when $r_s/r_p \sim 5$ this factor would reduce to $0.43$ if
$v_{rec}/v_{sw}$ remained the same.  Being less than or of order
unity, it highlights that despite having a collection area 100 times
larger, a robust magnetosphere collects less total ions at $r_s$ than
directly impact the atmosphere of an unshielded planet.  The fact that
$Q$ is not $<<1$ implies that the fate of ions trapped at $r_{s}$ may
be an important factor in evaluating the effectiveness of
magnetospheric shielding of atmospheres.  We should consider the
concentration of all of this collected flux onto polar caps covering a
surface area fraction $A_{cap}$ at the radius from the planet's center
where the ions are stopped, then the total wind flux impinging onto
these caps would be $q_{cap}= Q/A_{cap}$.  Today $q_{cap}$ is $>> 1$
because $A_{cap} <<1$. However, we recall that observations indicate
that atmospheric loss from Earth is not much greater than that at Mars
(which presently lacks an internal dynamo), contrary to that predicted
if the focusing effect was the dominant factor in atmospheric
loss. One reason for limited loss may be the return of ions in the
magnetotail. It may be that increases in focusing at the polar cap
associated with a magnetosphere are balanced by increased trapping in
the magnetotail. For the early Earth, we cannot assume $A_{cap} <<1$,
as discussed below.

\subsection{Implications for atmospheric escape at 3.4 to 3.45 Ga} 

\noindent

While the Paleoarchean geodynamo produced a magnetic field that
probably prevented whole-scale removal, magnetic field and solar wind
strengths suggest important modifications of the atmosphere by thermal
and non-thermal escape processes during the first billion years of
Earth evolution.  The $\sim$6 times greater XUV flux relative to today
would have heated the exosphere to many thousand Kelvin, promoting
thermal loss of hydrogen. The reduced standoff conditions would allow
greater interaction of solar wind protons with the extended
atmosphere, exacerbating thermal loss with non-thermal loss mechanisms
(Table 1).  The reduced standoff conditions would also be associated
with expansion of the polar cap, that area where open field lines
allow access of the solar wind to the deeper atmosphere.
\citet{Tarduno2010} used the latitude of the aurora to estimate the
polar cap area derived from a scaling law of Siscoe and Chen
\citep{Siscoe1975}:

\begin{equation}
{\rm cos} \lambda_{p} = \left( \frac{M_{E}}{M_{E_{0}}} \right)^{-1/6} P^{1/12} {\rm cos} \lambda_{p_{0}}
\end{equation}

\noindent
where $\lambda_{p}$ is the magnetic latitude of the polar cap edge,
$\lambda_{p_{0}}$ is the present-day value of 71.9$^{o}$, $M_{E_{0}}$
is the present-day dipole moment, and $P$ is the solar wind dynamic
pressure normalized to the present-day value ($\sim$2 nPa).  Using
this scaling law, \citet{Tarduno2010} suggested the polar cap could
have increased by as much as a factor of 3 under Paleoarchean solar
wind forcing.  In numerical simulations using the Space Weathering
Modeling Framework modules, \citet{Sterenborg2011} also found polar
cap enlargements, but by smaller amounts (15-50\%). We note that the
larger values reflected the most compressed magnetospheres (model B)
which represent much smaller standoff distances than those obtained in
the simulations. Equation 14 assumes a circular polar cap; day-night
side asymmetry should modify this shape, and the simulation results
show highly elliptical polar caps. The circular approximation may be a
better estimate of the open flux area sampled during one day.

The combined effect of enhanced XUV, reduced magnetic magnetopause
standoff, and increased polar cap area would have promoted the loss of
hydrogen and ultimately water from the Paleoarchean Earth. The
limiting factor for the water loss was probably not conditions of the
upper atmosphere, but the efficiency of water transport through a
lower atmosphere cold trap. However, high rates of photolysis related
to the large Paleoarchean UV flux may have enhanced the net removal of
water. Given that these reduced standoffs existed for hundreds of
millions of years (Figure 5), water transport and removal is
probable. This in turns implies that the Earth may have had a greater
water inventory at and prior to the Paleoarchean, allowing for
preservation of the modern oceans. The gradual removal of hydrogen
under reduced standoff conditions may also have been important for the
transformation of Earth's atmosphere from one that was mildly reducing
to one that was oxidizing.

The near certainty of extreme heating causing expansion of the
atmosphere prior to 3.45 Ga, and creating even greater opportunities
for atmospheric escape provides motivation for determining
solar-terrestrial interaction for even older times. We start with a
discussion of recent model predictions of dynamo onset, follow this
with estimates of solar winds for the first 700 million years of
Earth, and conclude with a discussion of the potential for magnetic
field strength observations for times $\gg$3.45 Ga.\\

\section{Delayed Dynamo Onset}

While the paleointensity values at 3.4 to 3.45 Ga are slightly less
than present-day, they are not remarkably different from variations
that may have occurred over the last 200 million years
\citep[e.g.][]{Aubert2010}. This raises the question of whether Earth
has always had a relatively strong magnetic field, dating from a time
shortly after core formation. Several diverse lines of reasoning have
been used to argue otherwise, with dynamo onset extending to times
just older than the 3.45 Ga constraint discussed above.  {\em
  Notwithstanding caveats discussed in section 4.4,} a delayed dynamo
onset and its associated long period without magnetic shielding imply
a correspondingly long episode of extreme water loss. In this case, a
huge initial water reservoir and/or large supply as a late veneer
\citep[e.g.][]{Albarede2009} may be needed to account for the
present-day conditions.

One of the most fascinating of these delayed dynamo hypotheses is
derived not from the study of terrestrial samples, but from
investigations of the Moon. Lunar ilmenite samples obtained during
Apollo missions have unusual nitrogen isotopic values; these have been
interpreted by \citet{Ozima2005} as reflecting nitrogen picked up from
Earth's atmosphere by the ancient solar wind and transported to the
lunar surface. An example of this process is the pickup of elements
from Venus observed today. The relevant point for our consideration is
that \citet{Ozima2005} recognized the importance of magnetic shielding
and hypothesized that the Earth's magnetic field was null (or of very
low intensity) during the time of nitrogen transfer, which was
constrained by the age of the Apollo samples (3.9 to 3.8 Ga). On the
basis of terrestrial constraints, this remains a viable hypothesis.

However, emerging paleointensity from lunar samples suggest an
important component of magnetic shielding may have been present on the
Moon.  The Moon appears to have a core that is at least partially
molten \citep{Weber2011} and magnetized crust \citep{Carley2012}, and
it has long been suspected that an ancient dynamo was once present on
the basis of analyses of Apollo samples \citep{Fuller1987}.  The
fidelity of paleointensity data from lunar samples is subject to
similar domain state constraints as discussed earlier for terrestrial
samples, but these limitations pale in comparison to greater
obstacles: the inherent thermal instability of FeNi phases that makeup
typical lunar magnetic grain populations and the effects of
impact-induced shock \citep{Fuller1974}.  Because of the thermal
instability, alternating field demagnetization and normalization with
applied fields has generally been used to estimate paleointensity
rather than Thellier analyses. Although there have been calibration
efforts \citep[e.g.][]{Gattacceca2004}, the accuracy of these data is
difficult to assess \citep[see][]{Lawrence2008} because the laboratory
method employed does not replicate the magnetization process.
Notwithstanding these uncertainties, recent studies have provided
evidence for lunar magnetizations $\sim$4.2 billion-years-old
\citep{Garrick-Bethell2009}, and possibly extending as young as 3.65
Ga \citep{Shea2012, Suavet2013}.  Beyond the issue of measurement
fidelity, the latter interpretations are somewhat controversial
because they call for a lunar field at times younger than the time
interval predicted for a viable dynamo powered by thermochemical
convection \citep{Stegman2003}. However, other dynamo mechanisms, such
as impact-stirring \citep{LeBars2011} and precession
\citep{Tilgner2005, Dwyer2011} might have powered a late lunar dynamo.
If confirmed, however, these late lunar magnetic fields pose a serious
challenge to the \citet{Ozima2005} hypothesis.  (To test his
hypothesis, \citet{Ozima2008} have proposed sampling the dark side of
the Moon.)

An entirely different approach to constraining the early dynamo
history is inspired by the detection of ultra-low seismic velocity
zones above the core, interpreted as dense melt lenses
\citep{Williams1996}.  \citet{Labrosse2007} interpreted these zones as
remnants of a once continuous dense melt layer. While present as a
continuous layer, \citet{Labrosse2007} postulated that it would form a
thermal boundary layer, limiting heat flow from the core and thus
suppressing thermal convection needed for dynamo generation. The dense
layer was thought to have dispersed into pods sometime between 4 and
3.4 Ga, after which a dynamo could be generated.

\citet{Aubert2010} expressed dynamo evolution in terms of scaling laws
derived from numerical simulations and present-day heat flow at the
core mantle boundary. In one model, a low CMB heat flow of 3 TW is
assumed. This model predicts the presence of an earliest Paleoarchean
and Hadean dynamo producing a field comparable in strength to today
(Figure 3). A deficiency of the model, however, is that is
underestimates modern field values. In another end member,
\citet{Aubert2010} considered a CMB heatflow similar of 11 TW; this is
closer to some estimates based on seismology \citep{Lay2008}. In this
model, dynamo onset is again delayed to times between 4 and 3.5 Ga.

The \citet{Labrosse2007} and \citet{Aubert2010} high CMB heatflow
models are just two in a class of thermal models that suggest the
onset of the dynamo may have been delayed because of mantle conditions
\citep{Jackson2013}: the mantle controls heat flow from the core and
if the lower mantle is too hot the heat flow will be limited (and the
core could theoretically heat up as has been proposed for small
bodies). The problem for early dynamo generation has been exacerbated
by recent changes in ideas on core thermal conductivity
\cite[e.g.][]{Olson2013}. Hence, onset of the dynamo is naturally
linked with the thermal regime of the lowermost mantle, and finding
constraints on the dynamo implicitly tells us about mantle history. We
will return to the potential for obtaining this record, but first we
address extending the solar wind to times older than 3.45 Ga.\\

\section{Solar Wind before 3.45 Ga}

In the prior considerations (equations 6-7), mass loss calculations
were not extended to the first $\sim$700 million years of Earth
history.  If we were to extend these loss rates to earliest times, it
would imply that the Sun was at least a few percent more massive than
otherwise assumed in standard models of solar evolution. A greater
luminosity associated with this more massive Sun could provide a
solution to the Faint Young Sun Paradox \citep[e.g.][]{Sackmann2003}.

However, while there are only a few solar analogs for times older than
3.45 Ga whose mass loss is constrained by H {\small I} Lyman-$\alpha$
observations \citep[e.g.][]{Wood2005,Wood2014}, the few that are
available seem to define mass loss rates that deviate from the trend
defined by older stars.  This suggests that there may be a different
magnetic topology affecting mass loss.  Specifically, the surfaces of
very young, active solar-like stars are thought to be dominated by
closed magnetic flux tubes \citep[e.g.][]{Schrijver2002}, whereas mass
loss mainly proceeds through open flux tubes
\citep[e.g.][]{Vidotto2009}.  We thus proceed assuming that the few
stellar analogs available are suggesting a different solar evolution
that must be addressed.

\citet{Suzuki2012} proposed a model which addresses this issue,
offering a scenario where the distribution of open magnetic flux tubes
important for mass loss for solar-like star evolves with time. Closed
flux loops dominate in the youngest times, limiting mass loss even
though stellar magnetic field intensity is high. Later, open flux
tubes occupy a greater portion of the solar surface. Eventually the
mass loss decreases with stellar magnetic field intensity
decrease. \citet{Suzuki2013} quantified this scenario as follows:

\begin{equation}
\dot{m} = \dot{M_{\odot}} \left(\frac{c_{M}}{0.023}\right) \left(\frac{R}{R_{\odot}}\right)^{3} \left( \frac{M}{M_{\odot}} \right)^{-1} \left(\frac{\rho_{o}}{10^{-7} \rm{g}\, \rm{cm}^{-3}}\right) \left (\frac{B_{r,0}f_{0}}{1.25\,\rm{G}}\right) \langle \left( \frac{ \delta v_{0}}{1.34\, \rm{km}\, \rm{s}^{-1}} \right)^{2}\rangle
\end{equation}

\noindent where $\dot{M}_{\odot}$ is the modern solar mass loss rate
via wind ($\sim$2 $\times$ 10$^{-14}$ M$_{\odot}$\,yr$^{-1}$), $c_{M}$
is a conversion factor (numerical simulations suggest $\sim$0.02), $R$
is the stellar radius, $M$ is the stellar mass, $\rho_{o}$ is density
(i.e. at the photosphere), $B_{r,0}$ is radial magnetic field, $f_{0}$
is the open flux tube filling factor over the photosphere, and $\delta
v_{0}$ is a velocity perturbation (which is expected to be a faction
of the sound speed at the photosphere).

 \citet{Suzuki2013} ran a series of magnetohydrodynamical simulations
 of stellar mass loss, varying the input parameters previously
 discussed over ranges of plausible values. They fit a power-law to
 their simulation results for X-ray surface flux $f_X$ $<$ 10$^{6}$
 erg\,cm$^{-2}$\,s$^{-1}$, in the regime where the kinetic energy of
 the winds are unsaturated. This X-ray surface flux $f_X$ corresponds
 approximately to that of the Sun at age $\sim$0.5 Gyr ($\sim$4.1 Ga),
 as estimated using the rotational and X-ray evolution relations of
 \citet{Mamajek2008}.

We adopt the power-law fits from \citet[][; their Equation
  25]{Suzuki2013}; and scale to an adopted mean modern solar soft
X-ray surface flux of $f_{X,\odot}$ = 3.7 $\times$ 10$^{4}$
erg\,cm$^{-2}$\,s$^{-1}$ \citep[based on discussion in][we adopt a
  modern-day solar luminosity of L$_{\rm X}$ $\simeq$ 10$^{27.35}$
  erg\,s$^{-1}$]{Judge2003}, to estimate mass loss and ram pressure
as:

\begin{equation}
\dot{m} = \dot{M}_{\odot} \left(\frac{R}{R_{\odot}}\right)^2 \left(\frac{F_{X}}{3.7\times 10^{4}\,\rm{erg}\,\rm{cm}^{-2}\,\rm{s^{-1}}}\right)^{0.82}
\end{equation}

\begin{equation}
P_{\rm SW \odot} = ({\rm 2 nPa}) \left(\frac{R}{R_{\odot}}\right)^2 \left(\frac{F_{X}}{3.7\times 10^{4}\,\rm{erg}\,\rm{cm}^{-2}\,\rm{s^{-1}}}\right)^{0.70}
\end{equation}

\noindent where again the modern average solar mass loss is
$\dot{M}_{\odot}$ $\sim$ 2 $\times$ 10$^{-14}$ M$_{\odot}$\,yr$^{-1}$.
The pressure P$_{SW}$ is evaluated at 1 AU, where we scale it to a
mean solar wind ram pressure of 2 nPa (based on four decades of
measurements compiled at \\ http://omniweb.gsfc.nasa.gov/html and
commensurate with a mean solar wind density of 7 cm$^{-3}$ and
velocity of 440 km\,s$^{-1}$.  Taking into account the expansion of
the Sun over its main sequence evolution, and its empirically
constrained rotational braking and X-ray luminosity evolution
\citep{Mamajek2008} we use these scaling relations based on
\citet{Suzuki2013} to estimate the mean solar mass loss and solar wind
pressure at 1 AU as a function of age.

At age 0.5 Gyr ($\sim$4.1 Ga), these relations predict both a solar
mass loss ($\sim$10$^{-12.7}$ M$_{\odot}$ yr$^{-1}$) and solar wind
pressure at 1 AU ($\sim$17 nPa) enhanced over current mean values by
only an order of magnitude (Model C, Figure 7; {\em see also
  Supplementary Content for further description}). A limitation of
this model is that it somewhat underestimates the present-day standoff
\citep[Model C: $\sim$9.5\,R$_{E}$ vs.  present-day value of
  10.1\,R$_{E}$;][]{Shue1997}.

Moreover, it should be emphasized that priority should be given to
additional observations as mass loss during the first 500 million
years of Earth history is constrained by analogy with data from only
the following 4 stars: $\pi^1$ UMa, $\xi$\,Boo A, Proxima Centauri,
and EV Lac \citep{Wood2014}. Of these, Prox Cen and EV Lac are active
M dwarfs, and $\xi$ Boo is part of a binary system where it is not
possible to accurately determine the relative contribution of winds
from the components. Thus far, there seems to be a ``Wind Dividing
Line'' for stars with X-ray surface flux $>$10$^6$
ergs\,cm$^{-2}$\,s$^{-1}$ \citep{Wood2014}, suggesting that our Sun's
mass loss during the main sequence stage may have peaked at
$\sim$10$^{-12}$ M$_{\odot}$\,yr$^{-1}$ near age $\sim$500 Myr (4.1
Gya), and been surprisingly lower (only $\sim$0.5-10$\times$ current
$\dot{M_{\odot}}$) at earlier ages.

Finally, we note that the above calculations have assumed that the
relevant pressure associated with inertia of the solar wind is
dominated by ram pressure with the solar wind magnetic pressure being
subdominant.  This is a good approximation for the present sun but a
faster rotation and stronger field at the surface of the young sun
could increase importance of the magnetic contribution to the solar
wind pressure. To see this, note that the strength of toroidal
magnetic field that arises from winding up of the poloidal field is
given by
\begin{equation}
B_{IMF} = B (r) =  B_{0} \frac{\omega}{v_{sw}}  \frac{r_{0}^{2}}{r}
\label{btor}
\end{equation}
\noindent
where $r$ is distance from Sun, $r_{0}$ is solar radius, $B_{0}$ is
open mean field at solar surface, $\omega$ is solar rotation speed
(faster for young sun) $v_{sw}$ is solar wind speed.  If we ignore the
thermal pressure, the total solar wind pressure at a given distance
from the sun would then be the sum of the ram + magnetic pressure at
that distance, and is then given by
\begin{equation}
\rho_{sw} v_{sw}^2 \left(1
+ {B_{IMF}^2/ 8\pi \over   \rho_{sw} v_{sw}^2}\right) =
\rho_{sw} v_{sw}^2 \left[1 + {B_0^2 \omega^2  r_0^4\over 8\pi  \rho_{sw} v_{sw}^4 r^2} \right],
\label{btor2}
\end{equation}
using equation 18.  Using the numerical values appropriate for the
present sun of $B_0 =2{\rm G}$, $\omega = 3 \times 10^{-6} {\rm
  rad/s}$, $r_0= 7 \times 10^{10} {\rm cm}$, $r =1.5 \times 10^{13}
{\rm cm}$, $\rho_{sw} = 10^{-23} {\rm g/cm^3}$, and $ v_{sw} = 4
\times 10^{7} {\rm cm/s}$, the magnetic correction term in parenthesis
of equation 18 is $0.006$ today.  If however $\omega$ and $B_0$ were
significantly larger at early times compared to the increase in
$\rho_{sw} v_{sw}^{4}$ at early times, then the magnetic correction
term could increase.  But depending on the mechanism that drives the
solar wind, $\rho_{sw} v_{sw}^{4}$ itself could depend on $B_0$ and
$\omega$ so it is also possible that this correct term does not
substantially increase.  A detailed discussion of these subtleties are
beyond the present scope.

\section{External source of field for a  planet lacking a core dynamo}

Given the increased solar wind pressure associated with the young Sun,
it is prudent to revisit the question of solar-induced magnetic
fields. This is relevant for gauging the minimum field that might be
expected in the absence of a early internally-generated dynamo
magnetic field for Earth. A lower limit for the external magnetic
field is that due to the solar wind itself, and is given by equation
18 which suggests a value between about 7 and 10 nT for Earth today.
However, when the supersonic solar wind impacts the atmosphere, field
lines are compressed, and an enhancement of the field is expected.
The maximum field strength would correspond to the case in which all
of the wind ram pressure is converted into magnetic pressure, that is
\begin{equation}
\frac{B^{2}}{8\pi} \sim \rho_{sw} v_{sw}{^2}
\end{equation}

\noindent
where $\rho_{sw}$ is the solar wind density. Such equipartition would
give a maximum field of about 60 nT today (if Earth lacked an internal
magnetic field). This field estimate thus scales with the square root
of the solar wind ram pressure. For solar wind pressures of 100 to
1000 times those of today this estimate of the external field would
give 0.6 to 2 $\mu$T.  We note that for Model C, these values are
approached in steady state only for times $>$4.5 Ga.

The field will would begin to be amplified near the exobase as field
lines are compressed, and deeper in the atmosphere as the field lines
slip through atmospheric material.  The amplified field lines will
drape around Earth from the force for the continuous solar wind.  The
mixing may also be contemporaneous with turbulence that produces
complex directional changes in the field on small scales. However,
because the solar wind properties vary on such a large scale compared
to that of the distance from exobase to Earth's surface, the overall
magnitude of the supply energy and spatially averaged field are not
expected to vary greatly over the daylight side of the planet.  Thus,
while there will be overall magnitude variability on day time-scales,
a small external field could be sensed by a slowly cooled magnetic
mineral on Earth's surface. This serves as a bound on what
paleomagnetic studies might be able to resolve about the earliest core
dynamo. We also note that this external field could also serve as a
seed-field for the start-up of a core dynamo if it penetrates all the
way to the core. \\

\section{Paleoarchean and Hadean Magnetic Sands\\}

Are there rocks that can be sampled to extend the paleointensity
record back in time, before 3.45 Ga?  The Pilbara Craton of Western
Australia hosts Paleoarchean greenshist facies rocks, but these can
extend the record from Barberton by only a few tens-of-millions of
years. The key time interval for testing the geodynamo
presence/absence question is considerably older (cf. Figure 3).

Unfortunately, other terrestrial rocks with ages $\gg$3.45 Ga have
been multiply deformed and metamorphosed to amphibolite-facies (or
higher); the thermal and associated chemical transformations remove
these from further paleomagnetic considerations. This includes the
oldest known rocks, including parts of the 4.03 Ga Acasta Gneiss of
northwestern Canada \citep{Bowring1999}.

However, there is another recorder: silicate crystals hosting magnetic
inclusions that have eroded from primary igneous rocks and are now
found in younger sedimentary units. The benefit of this approach is
that it might allow us to sample time intervals that are not otherwise
available because the original igneous rocks have been lost to
erosion. This requires dating of the silicate crystal itself,
something that is commonly done with some sedimentary components
(e.g. zircons).  We note that in using this approach paleolatitudes
will typically not be available, because as sedimentary detritus, the
orientation of a given silicate crystal relative to horizontal is
lost. In this case, inferences on the nature of the global field
strength will be limited to the factor of 2 variation of a dipole with
latitude. Nevertheless, this is sufficient to test the
presence/absence of the geodynamo.  We describe approaches to the
paleointensity investigation of silicates found in younger sedimentary
units below, focusing on what is arguably the most famous unit, the
Jack Hills metaconglomerate.\\

\subsection{Jack Hills Metaconglomerate}

The Jack Hills metaconglomerate is located on the northern edge of the
Yilgarn Craton of Western Australia. The unit is rich in zircons and
in places, most notable the ``Discovery outcrop", up to 12\% of the
zircons have ages of 4 billion years or older
\citep{Compston1986}. The oldest zircons are dated to $\sim$4.4 Ga,
only some 150 million years after formation of the planet
\citep{Wilde2001}.  The observation of elevated $\delta^{18}$O in some
zircons indicates the incorporation of hydrated rocks in magmas from
which they grew.  This in turn suggests that Earth had oceans at this
time \citep[e.g.][]{Cavosie2007}. Because water is an essential
ingredient for life as we know it, this observation has motivated some
to suggest that the Archean/Hadean boundary should be moved from the
traditional age of 3.9 Ga to 4.2 Ga \citep{Valley2006}.  Concerned
efforts to find the source of the zircons have failed, leading to the
conclusion that these have been lost to erosion, possibly during
intense chemical weathering, as supported by $\delta^{7}$Li data
\citep{Ushikubo2008}.

The Jack Hills sediments have experienced peak metamorphism of 420 to
475 $^{o}$C \citep{Rasmussen2010} and then have been deformed
\citep{Spaggiari2007a, Spaggiari2007b} into a stretched pebble
conglomerate. As in the studies on the 3.45 Ga dacites from the
Barberton Greenstone Belt, a first step in assessing the viability of
the Jack Hills sediments is to conduct a paleomagnetic conglomerate
test.

To address the deformation issue, \citet{Tarduno2013} sampled the
interiors of cobble-sized rounded clasts, following the idea that they
may have primarily undergone rolling motion rather than stretching,
limiting penetrative deformation. Magnetic susceptibility versus
temperature data indicate the presence of the Verwey transition, the
cubic to monoclinic change in the crystal structure of magnetic seen
on cooling through 120 K \citep{Verwey1939}. Thermal demagnetization
reveal a complex series of magnetizations removed at low to
intermediate temperatures. Some of these magnetizations are grouped,
indicating the potential of magnetization during one of the Archean to
Proterozoic metamorphic events that affected the Jack Hills. At
relatively high unblocking temperatures, often greater than 550
$^{o}$C, a distinct component trending to the origin of orthogonal
vector plots of the magnetization is commonly observed. This component
is scattered, and statistical tests indicate formal passage of the
conglomerate test (Figure 8).

This indicates magnetization of the interior part of the cobble only
prior to the deposition of the conglomerate; unlike the case of the
dacite conglomerate of the Barberton Greenstone Belt, a substantial
time period elapsed between the deposition of the conglomerate and the
potential age of the mineral components. Most of the Jack Hills
sediments appear to have been deposited at ca. 3 Ga, but rare younger
units (as young as 1.2 Ga) have been tectonically interleaved.
Notwithstanding the potential for some Proterozoic ages associated
with these tectonic slivers, it appears that select components of the
Jack Hills metasediments have the potential to preserve magnetizations
at least as old as ca. 3 Ga.\\

\subsection{Quartz and Zircon carriers}

Is there a 4 billion-year-old (or older) rock preserved as a clast in
the Jack Hills metaconglomerate?  The intense chemical weathering
called upon for the Hadean would seem to suggest this is unlikely, but
it more possible that individual quartz grains composing the Jack
Hills clasts are of this age. As discussed above, these quartz grains
commonly contain magnetite, and thus they are viable paleomagnetic
recorders. However, determining the age of these quartz grains is
difficult. In the Jack Hills metaconglomerates, quartz grains often
contain zircon inclusions; hence these could provide some age
constraints.

Zircon having magnetic inclusions is itself a potential paleomagnetic
recorder (hinted at in \cite{Dunlop1997} and discussed explicitly in
\cite{Tarduno2009} and \cite{Nelson2010}), although one that
challenges paleomagnetic measurement technology because of their small
size (typically $<$300 microns) and related small number of magnetic
included particles.  However, high resolution 3-component DC SQUID
magnetometers have been used to detect paleointensities from single
zircon crystals using Thellier-Coe techniques \citep{Cottrell2013}.
It should also be kept in mind that the thermal history of these
grains from their formation to their final incorporation into a
conglomerate, may not be straightforward. Radiometric age dating
coupled with petrologic examinations (i.e. the presence of
overgrowths) has the potential to distinguish zircons which may record
paleomagnetic signals dating from their initial formations to those
that have been reset due to reheating (at times much older than the
conglomerate age).

While retrieving a paleomagnetic history from such zircons is a goal,
it should be remembered that the complications of zircon resetting are
somewhat less important when viewed in the context of simple field
absence/presence criteria. That is, the possibility of a ``false
positive" identification of the geomagnetic field for times older than
3.45 Ga is omipresent because of subsequent geological activity. But
the identification of an anomalously low or null field (within the
bounds of recording zero discussed above), if this existed before 3.45
Ga, would be otherwise difficult to explain because of the abundance
of opportunities for later magnetization.\\

\section{Summary and Future Potential}

The ``habitable zone" is classically defined as that distance from a
star where liquid water can exist. Even given birth of a planet in
this zone, there is no assurance that a habitable planet will evolve
given the potential for water loss. The magnetic field is a key factor
that must be considered in determining whether a terrestrial-like
planet will retain its water. The preservation potential will in turn
depend on the balance of stellar wind pressure and magnetic field
strength.  Stellar wind history will be a function of star spin rate
and stellar evolution.  For terrestrial-like planets, salient
variables include the time of onset and duration of the dynamo (which
are related to the efficiency of heat removal from the core),
especially during the first billion years after planet formation.

The magnetic field has competing effects with respect to atmospheric
retention (and ultimately water survival). Understanding better the
net influence of these effects is itself an important direction of
future research. For example, an increased magnetic field provides
more pressure to abate the solar wind dynamic pressure and increase
the magnetopause radius but the larger magnetopause also means a
larger collecting area for solar wind flux during phases of magnetic
reconnection.  In addition, a strong ordered dipole magnetic field may
provide a pathway for concentrating the solar wind flux toward polar
caps where its local mass-loss effects may be exacerbated. Yet, this
same ordered field provides the magnetic topology for recapturing this
mass in the opposite hemisphere such that the net global atmospheric
mass loss might not be affected.  Comparing similarities and
differences of mass loss from planets within our present day solar
system will be helpful for progress.  Presently available data support
the net atmospheric protective role of dynamo magnetic fields.

Ultimately, understanding differences between present-day and early
solar wind conditions and influences on Earth's atmosphere (and the
atmospheres of other terrestrial planets) is an important goal. To
constrain early solar winds, additional observations on solar analog
stars younger than 500 Myr (4.1 Ga) are needed.  For the Paleoarchean
Earth (at ca. 3.4-3.45 Ga), the balance between core dynamo field
values and increased solar wind pressure results in standoff of the
solar wind to distances half of those of the present-day, suggesting
some atmospheric loss and removal of water.  This erosive potential
implies a more water-rich initial Earth, and/or the delivery of water
as a late veneer, to account for the present terrestrial reservoir.
The need for large early water reservoirs, or late replenishment, is
exacerbated if onset of the geodynamo was delayed by a hot lower
mantle.  Therefore, from a deep Earth perspective, better constraints
on Hadean lower mantle evolution and its interplay with core heat loss
could aid our understanding of the geodynamo/shielding history.  These
linkages between the early mantle, core dynamo and atmospheric
development are equally relevant for Mars (and possibly Venus).  From
an observational perspective on Earth, there is potential in early
Archean and Hadean zircons (and other sedimentary crystals) hosting
magnetic inclusions to record the first billion years of the geodynamo
using single crystal paleointensity methods. This remains a grand
challenge, requiring the most sensitive magnetometers and the
development of methods to understand the effects of geologic history
after zircon formation. Many of these ongoing instrument and technique
developments will have continued application in the analysis of
Martian rocks retrieved by future sample return missions.\\[5mm]

\section{Acknowledgements}
We thank Rory Cottrell, Richard Bono, Axel Hofmann, Ariel Anbar and
David Sibeck for helpful conversations, and the editor and two
anonymous reviewers for their comments. EGB acknowledges partial
support from the Simons Foundation.This work was supported by NSF.

\bibliography{tarduno2_biblio}

\newpage

\noindent
\textbf{Supplementary Content}

\bigskip

\noindent
{\it Reconstructing the Past Sun}

We combine a theoretical stellar evolutionary track for a 1
M$_{\odot}$ star with observational constraints on the time-evolution
of various stellar parameters for Sun-like stars to produce a
reconstruction of the Sun's characteristics at geologically and
astronomically relevant times ({\bf Table 2}). The solar parameters
are listed at the starts of various geological time periods, adopting
recent ages from the Geological Society of America \citep{Walker13}.
We adopt the 1 M$_{\odot}$ stellar evolutionary track from
\citet{Bressan12}, which employs the recent \citet{Caffau11} mixture
for protosolar chemical composition. We make minor systematic shifts
to the luminosity and effective temperature of the evolutionary track
at the $\sim$1\%\, level in order to match the current Sun at an
adopted meteoritic age for the solar system \citep[4568
  Myr;][]{Bouvier10,Amelin10}. We adopt the revised stellar parameters
compiled by \citet{Mamajek12}: effective temperature T$_{\rm eff}$ =
5772 K, luminosity $L$ = 3.827 $\times$ 10$^{33}$ erg\,s$^{-1}$, and
radius $R$ = 695660 km.  Spectral types were estimated through the new
main sequence effective temperature scale of \citet{Pecaut13}
(although in practice spectral types for G-stars are rarely quoted to
better than $\pm$1 subtype precision). The X-ray luminosity evolution
as a function of rotation was calibrated to the data from
\citet{Wright11}, but was adjusted to pass through the Sun's current
combination of average X-ray luminosity and rotation period
\citep[parameterized via Rossby number;][]{Mamajek08}. We have
included an estimate of the mean emission-measure-averaged coronal
temperature log\,$\tilde{T}_X$ as a function of the Sun's age, based
on a custom fit to the data of \citet{Telleschi05} and the modern Sun
\citep{Peres00} as a function of mean X-ray luminosity in the ROSAT
X-ray bandpass (L$_X$): log\,$\tilde{T}_X$ $\simeq$ -1.54 +
0.282\,log\,L$_X$ (L$_X$ in erg\,s$^{-1}$).

We estimate the Sun's current average solar wind mass loss as
follows. Based on $\sim$15,000 daily solar wind measurements in the
OmniWeb database\footnote{Goddard Space Flight Center Space Physics
  Data Facility: http://omniweb.gsfc.nasa.gov/.} between 1963 and
2014, we estimate a mean daily solar wind density of $n$ = 6.94
cm$^{-3}$ and mean solar wind velocity of $V_{SW}$ = 439
km\,s$^{-1}$. Extrapolating these values over 4$\pi$ steradians, one
would estimate the solar wind mass loss rate to be
1.94$\times$10$^{-14}$ M$_{\odot}$\,yr$^{-1}$. Results from the
Ulysses mission \citep{Goldstein96} show that at high heliographic
latitude ($>$20$^{\circ}$) the solar wind has a product of density and
velocity approximately half that at lower latitudes. We take this
result into account and multiple our original estimate by $\sim$2/3,
leading to a spherically-averaged mean solar mass loss via solar wind
of $\dot{M}_{\odot}$ = 1.3\,$\times$\,10$^{-14}$
M$_{\odot}$\,yr$^{-1}$.

Three mass loss rates are listed in Table 2: $\dot{M}_{W}$ is
estimated following the observational trends from \citet{Wood14},
$\dot{M}_{S}$ is estimated following the simulations of
\citet{Suzuki13}, and $\dot{M}_{CME}$ is the estimated mass loss due
solely to coronal mass ejections from flares
\citep{Drake13}. Magnetopause radii are estimated following
\citet{Tarduno10}, assuming Earth magnetic field strengths equal to
the current value (R$_{S,1}$) and half (R$_{S,1/2}$; similar to that
measured for 3.45 Gya by Tarduno et al. 2010). Interplanetary magnetic
field pressure was assumed to be negligible at all periods compared to
the dynamical ram pressure of the wind.

Several of the stellar parameters for the Sun and Sun-like stars
(mostly related to magnetic activity) are observationally well
correlated with rotation period \citep[e.g.][]{Mamajek08}, so we made
a careful reassessment of the Sun's likely rotational evolution
through study of Sun-like main sequence stars of different ages. Based
upon a literature review of measured rotation rates of $\sim$1
M$_{\odot}$ ($\pm$10\%) stars in ten young star clusters\footnote{In
  age order: Pleiades, M50, M35, M34, M11, Coma Ber, M37, Praesepe,
  Hyades, and NGC 6811.}  and older field stars\footnote{In age order:
  18 Sco, Sun, $\alpha$ Cen A \& B (mean), 16 Cyg B.} we derive a
revised version of the Skumanich law \citep{Skumanich72}: P$_{rot}$ =
25.5($t/t_{\odot}$)$^{0.526\,\pm\,0.022}$ day, where $t$ is stellar
age, $t_{\odot}$ is the Sun's age (4568 Myr), and the relation is
empirically constrained between $\sim$0.1-7 Gyr.  A fit of Sun-like
stars in young clusters and (older) field stars {\it omitting} the
Sun, yields a nearly identical relation, {\it predicting} the modern
Sun's rotation period to be 25.4 day. We surmise that the Sun is a
normal rotator (within $\pm$1 day) for its mass and age.

Has mass loss via the solar wind had an impact on the Sun's luminosity
evolution since reaching the main sequence? An enhanced early solar
wind has been proposed to be a potential solution to the Faint Young
Sun paradox \citep[e.g.][]{Sackmann03}. We have surveyed the recent
literature for published mass loss estimates and trends for Sun-like
stars, as a function of age and/or X-ray luminosity, as well as
theoretical predictions
\citep[e.g.][]{Holzwarth07,Cranmer11,Drake13,Suzuki13,Wood14}.  Thus
far, these recent studies are consistent with a total solar main
sequence mass loss in the range $\sim$0.01-0.4\%.  In the same period,
the Sun has lost $\sim$0.03\%\, of its mass due to radiative losses
through converting mass to energy \citep{Sackmann03}. The
\citet{Bressan12} stellar evolutionary tracks are consistent with
having zero-age main sequence luminosities of L$_{ZAMS}$ $\simeq$ 0.70
(M/M$_{\odot}$)$^{4.535}$ L$_{\odot}$ for solar composition stars
within 10\%\, of a solar mass. After 4.6 Gyr, the total predicted
solar mass loss due to solar wind and radiative losses is in the range
$\sim$0.04-0.4\%, so the Sun could have plausibly been negligibly more
luminous ($\sim$0.2-1.8\%) early in its main sequence phase. Hence,
current observational and theoretical constraints on the mass loss
history of the Sun seem inconsistent with enhanced early stellar winds
providing a parsimonious solution to the Faint Young Sun paradox
\citep[e.g.][]{Sackmann03}.

\begin{sidewaystable}
\caption{Solar Parameters: Zero-Age Main Sequence to Present}
\footnotesize
 \begin{tabular}{ccclccccccccccl}
$\tau$ & Age & T$_{\rm eff}$ & Spec. & Lum.           & Rad. &
log\,R$_X$ & log\,L$_{X}$ & log\,\~{T}$_X$ & $\dot{M}_{W}$ &
$\dot{M}_{S}$ & $\dot{M}_{CME}$ & R$_{S,1}$ & R$_{S,1/2}$ & Name
of Starting Geological Period\\
Gya     & Gyr  & K                & Type & L/L$_{\odot}$ &
R/R$_{\odot}$ & dex & erg\,s$^{-1}$ & K & M$_{\odot}$ yr$^{-1}$ &
M$_{\odot}$ yr$^{-1}$ &  M$_{\odot}$ yr$^{-1}$ & R$_{Earth}$ &
R$_{Earth}$ & ...\\
 \hline \hline 
4.525 & 0.045 & 5630 & G5.4V & 0.686 & 0.871 & -3.33 & 30.1 & 6.96 & -12.6 & -11.5 & -10.1 & 4.7 & 3.7 & {\it Zero-Age Main Sequence (ZAMS)}\\
4.450 & 0.120 & 5645 & G5.2V & 0.707 & 0.879 & -3.92 & 29.5 & 6.79 & -13.4 & -11.9 & -10.9 & 5.5 & 4.4 & {\it Pleiades Cluster Age}\\
4.000 & 0.570 & 5660 & G5.0V & 0.735 & 0.891 & -4.85 & 28.6 & 6.54 & -12.0 & -12.7 & -12.3 & 7.0 & 5.6 & Archaen Eon/Eoarchean Era\\
3.920 & 0.650 & 5662 & G4.9V & 0.739 & 0.893 & -4.93 & 28.5 & 6.51 & -12.1 & -12.8 & -12.4 & 7.2 & 5.7 & {\it Hyades Cluster Age}\\
3.600 & 0.970 & 5672 & G4.4V & 0.756 & 0.900 & -5.18 & 28.3 & 6.45 & -12.4 & -13.0 & -12.7 & 7.7 & 6.1 & Paleoarchean Era\\
3.450 & 1.120 & 5676 & G4.2V & 0.764 & 0.904 & -5.27 & 28.2 & 6.42 & -12.5 & -13.0 & -12.9 & 7.8 & 6.2 & Barberton Greenstone Belt dacite\\
3.200 & 1.370 & 5684 & G3.9V & 0.777 & 0.909 & -5.40 & 28.1 & 6.39 & -12.7 & -13.1 & -13.1 & 8.1 & 6.4 & Mesoarchean Era\\
2.800 & 1.770 & 5696 & G3.6V & 0.800 & 0.918 & -5.56 & 27.9 & 6.35 & -12.9 & -13.2 & -13.4 & 8.4 & 6.7 & Neoarchean Era\\
2.500 & 2.070 & 5705 & G3.4V & 0.818 & 0.926 & -5.67 & 27.8 & 6.32 & -13.0 & -13.3 & -13.6 & 8.6 & 6.8 & Proterozoic Eon\\
2.300 & 2.270 & 5710 & G3.2V & 0.830 & 0.931 & -5.73 & 27.8 & 6.30 & -13.1 & -13.4 & -13.7 & 8.8 & 6.9 & Rhyacian Period\\
2.050 & 2.520 & 5718 & G3.1V & 0.846 & 0.937 & -5.81 & 27.7 & 6.28 & -13.2 & -13.4 & -13.8 & 8.9 & 7.1 & Orosirian Period\\
1.800 & 2.770 & 5725 & G2.9V & 0.862 & 0.944 & -5.87 & 27.6 & 6.27 & -13.3 & -13.5 & -13.9 & 9.0 & 7.2 & Stratherian Period\\
1.600 & 2.970 & 5731 & G2.8V & 0.876 & 0.949 & -5.92 & 27.6 & 6.25 & -13.3 & -13.5 & -14.0 & 9.1 & 7.2 & Mesoproterozoic Era\\
1.400 & 3.170 & 5736 & G2.7V & 0.890 & 0.955 & -5.97 & 27.6 & 6.24 & -13.4 & -13.5 & -14.1 & 9.2 & 7.3 & Ectasian Period\\
1.200 & 3.370 & 5742 & G2.6V & 0.904 & 0.961 & -6.01 & 27.5 & 6.23 & -13.4 & -13.6 & -14.2 & 9.3 & 7.4 & Stenian Period\\
1.000 & 3.570 & 5747 & G2.5V & 0.919 & 0.967 & -6.06 & 27.5 & 6.22 & -13.5 & -13.6 & -14.2 & 9.4 & 7.5 & Tonian Period\\
0.850 & 3.720 & 5751 & G2.4V & 0.930 & 0.971 & -6.08 & 27.5 & 6.22 & -13.5 & -13.6 & -14.3 & 9.4 & 7.5 & Cryogenian Period\\
0.635 & 3.935 & 5757 & G2.3V & 0.947 & 0.978 & -6.13 & 27.4 & 6.21 & -13.6 & -13.6 & -14.4 & 9.5 & 7.6 & Ediacaran Period\\
0.541 & 4.029 & 5759 & G2.2V & 0.954 & 0.981 & -6.15 & 27.4 & 6.20 & -13.6 & -13.6 & -14.4 & 9.6 & 7.6 & Cambrian Period\\
0.485 & 4.085 & 5760 & G2.2V & 0.959 & 0.983 & -6.15 & 27.4 & 6.20 & -13.6 & -13.7 & -14.4 & 9.6 & 7.6 & Ordovican Period\\
0.444 & 4.126 & 5762 & G2.2V & 0.962 & 0.985 & -6.17 & 27.4 & 6.20 & -13.6 & -13.7 & -14.4 & 9.6 & 7.6 & Silurian Period\\
0.419 & 4.151 & 5762 & G2.2V & 0.964 & 0.985 & -6.17 & 27.4 & 6.20 & -13.6 & -13.7 & -14.4 & 9.6 & 7.6 & Devonian Period\\
0.359 & 4.211 & 5764 & G2.2V & 0.969 & 0.987 & -6.18 & 27.4 & 6.20 & -13.6 & -13.7 & -14.5 & 9.6 & 7.6 & Carboniferous Period\\
0.299 & 4.271 & 5765 & G2.1V & 0.974 & 0.990 & -6.19 & 27.4 & 6.19 & -13.7 & -13.7 & -14.5 & 9.7 & 7.7 & Permian Period\\
0.252 & 4.318 & 5766 & G2.1V & 0.978 & 0.991 & -6.20 & 27.4 & 6.19 & -13.7 & -13.7 & -14.5 & 9.7 & 7.7 & Triassic Period\\
0.201 & 4.369 & 5767 & G2.1V & 0.983 & 0.993 & -6.21 & 27.4 & 6.19 & -13.7 & -13.7 & -14.5 & 9.7 & 7.7 & Jurassic Period\\
0.145 & 4.425 & 5769 & G2.1V & 0.988 & 0.995 & -6.22 & 27.4 & 6.19 & -13.7 & -13.7 & -14.5 & 9.7 & 7.7 & Cretaceous Period\\
0.066 & 4.504 & 5771 & G2.0V & 0.994 & 0.998 & -6.23 & 27.3 & 6.18 & -13.7 & -13.7 & -14.5 & 9.7 & 7.7 & Paleogene Period\\
0.023 & 4.547 & 5771 & G2.0V & 0.998 & 0.999 & -6.24 & 27.3 & 6.18 & -13.7 & -13.7 & -14.6 & 9.7 & 7.7 & Neogene Period\\
0.003 & 4.567 & 5772 & G2.0V & 1.000 & 1.000 & -6.24 & 27.3 & 6.18 & -13.7 & -13.7 & -14.6 & 9.7 & 7.7 & Quaternary Period\\
\hline \hline
\end{tabular}
\vspace{-7mm}
\end{sidewaystable}

\newpage
\noindent
\textbf{Figures}\\[2mm]

\noindent
\begin{figure}[!ht]
\centering
\includegraphics[width=\linewidth]{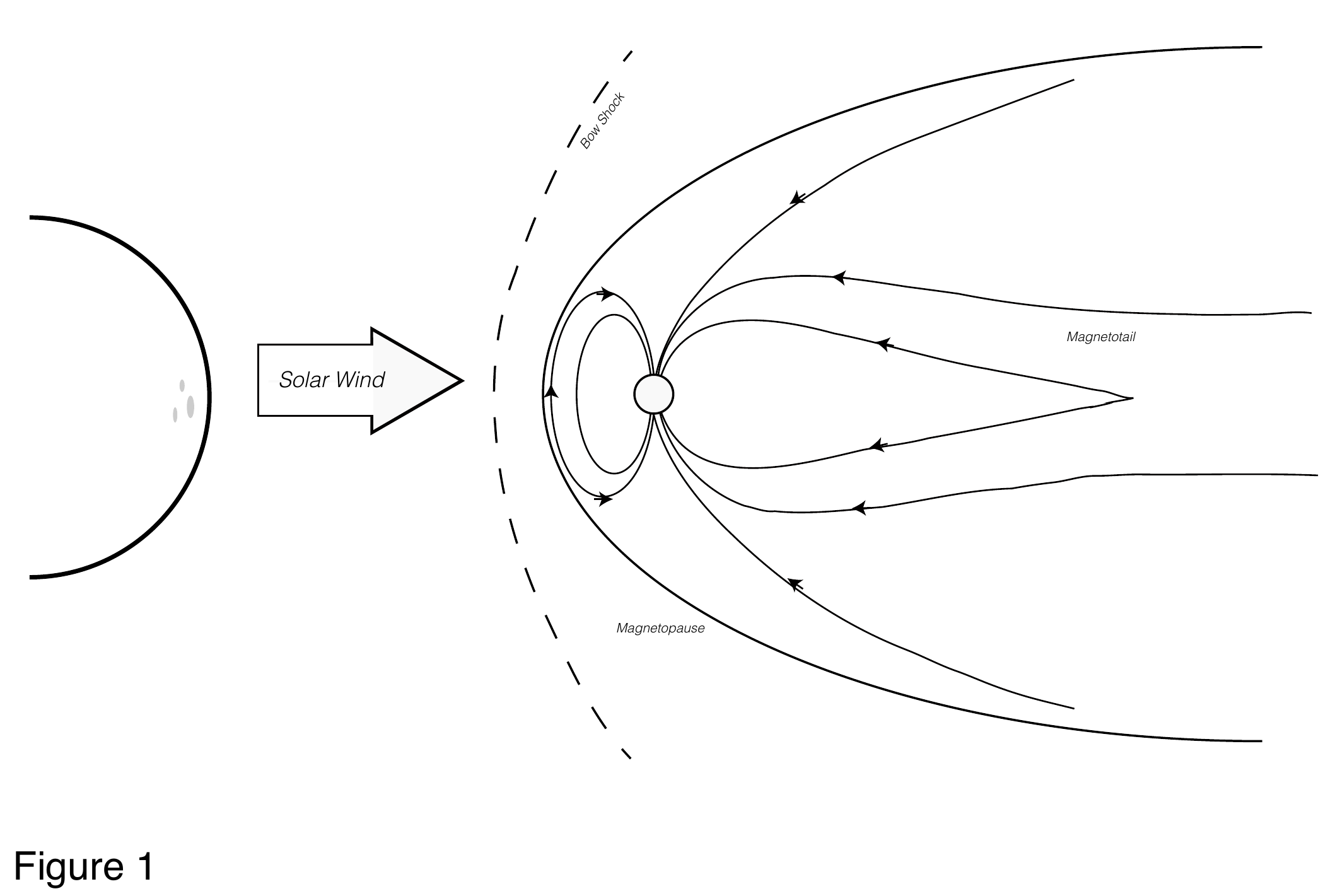}
\caption{
Magnetosphere (not to scale) shaped by the
interaction of the solar wind (with pressure $P_{sw}$ and Earth
magnetic field ($M_{E}$)). The point where the solar wind pressure is
balanced by the magnetic field is the standoff distance ($r_{s}$).}
\end{figure}

\noindent
\begin{figure}[!ht]
\centering
\includegraphics[width=\linewidth]{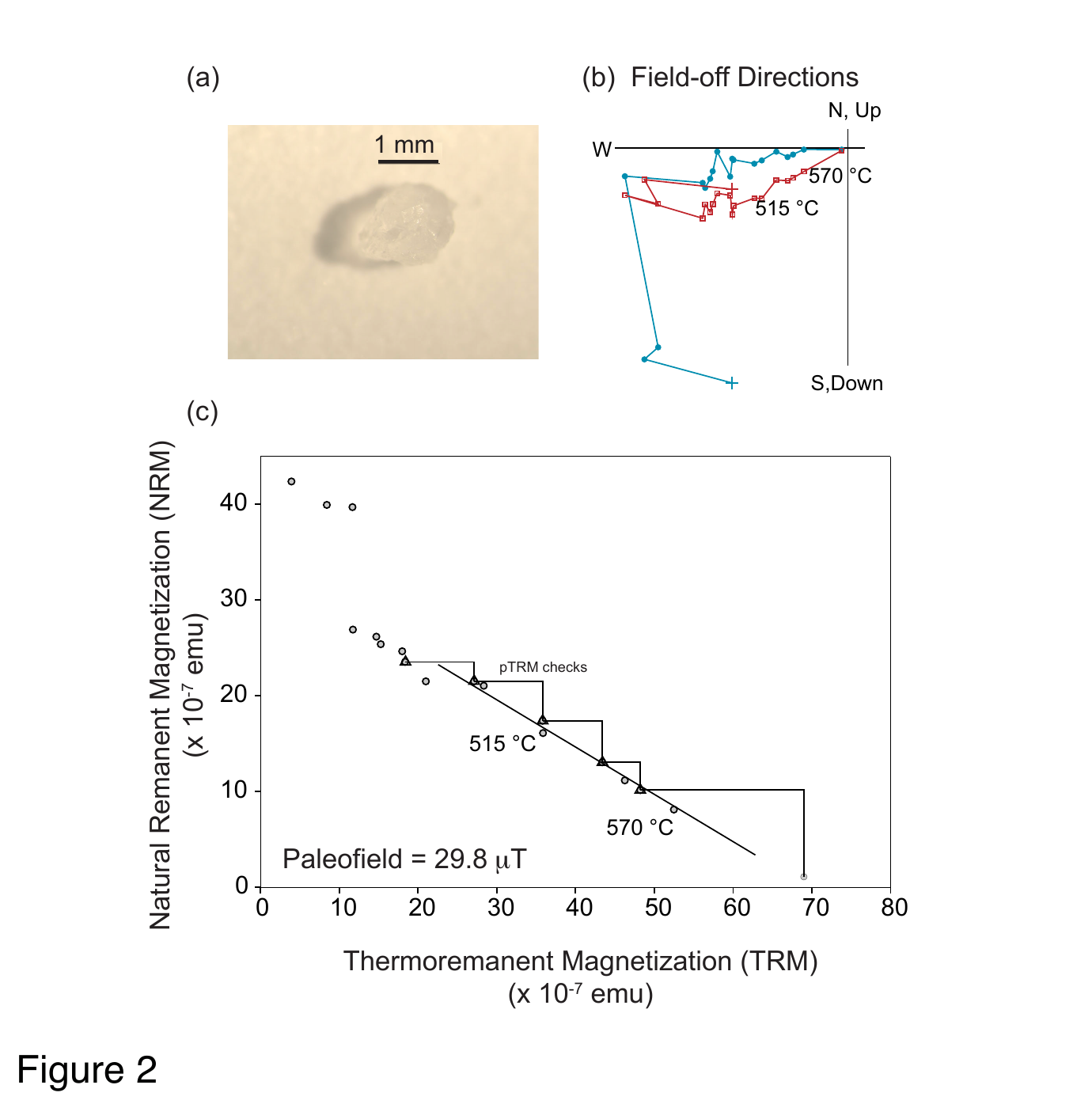}
\caption{Example Thellier-Coe paleointensity experiment
conducted on a single silicate crystal. a. Sample analyzed (quartz
phenocryst with magnetite inclusions). b. Orthogonal vector plot of
``field off" steps (see text). These define a component that trends
toward the origin between approximately 515 and 570 $^{o}$C. Red is
inclination, blue is declination (note absolute orientation is
arbitrary because the sample is unoriented). (c) Natural remanent
magnetization (NRM) lost versus thermoremanent magnetization gained
(TRM). TRM was acquired using an applied field of 60 $\mu$T. The slope
of the line, for the temperature range where field-off data show
linear decay to the origin (b) defines the paleofield
strength. Example from \citep{Tarduno2010}.}
\end{figure}

\noindent
\begin{figure}[!ht]
\centering
\includegraphics[width=\linewidth]{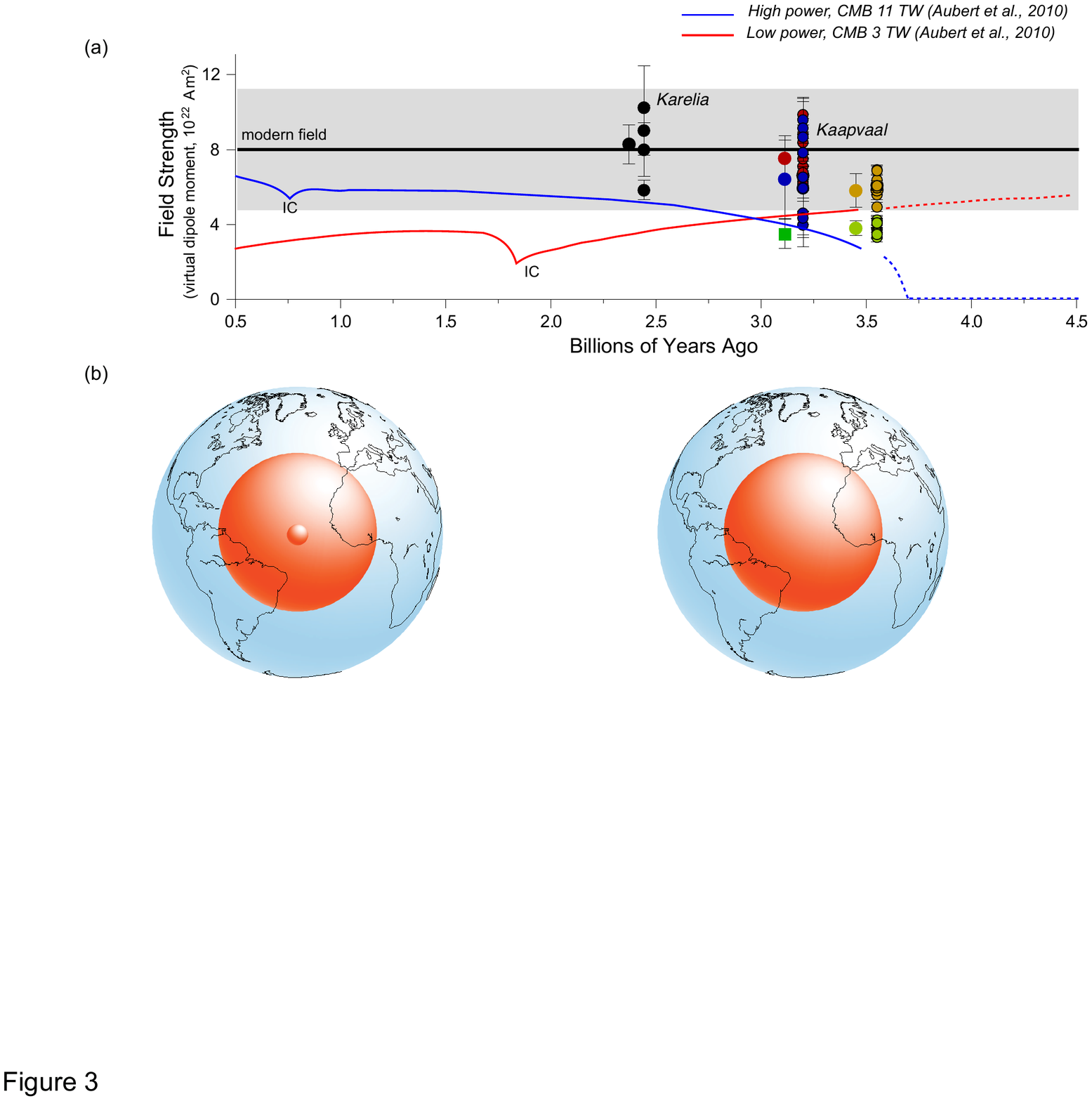}
\caption{
Archean paleointensity constraints based on single
silicate crystal paleointensity and inner core growth. a. Field
strength values are from Karelia (black, \citep{Smirnov2003} at
$\sim$2.5 Ga, the Kaapvaal Craton at 3.2 Ga (red, blue, circles
\citep{Tarduno2007} and 3.4-3.45 Ga (yellow, green, circles
\citep{Tarduno2010}).  Individual determinations (outlined circles)
for the Kaapvaal craton are shown with mean (to left, circles without
outlines). Green square is the 3.2 Ga data corrected for cooling
rate. As discussed in the text, this cooling rate correction probably
results in an underestimate of the true field value.  Also shown are
two models for field intensity history from
\citep{Aubert2010} based on high/low values of present-day core-mantle
boundary heat flow. IC is the time of onset of inner core growth for
the models. Many recent models invoke the onset of inner core growth
younger than 1 Ga (b). In nearly all models, a long-lived Paleoarchean
geodynamo would need to be entirely thermally driven.}
\end{figure}

\noindent
\begin{figure}[!ht]
\centering
\includegraphics[width=\linewidth]{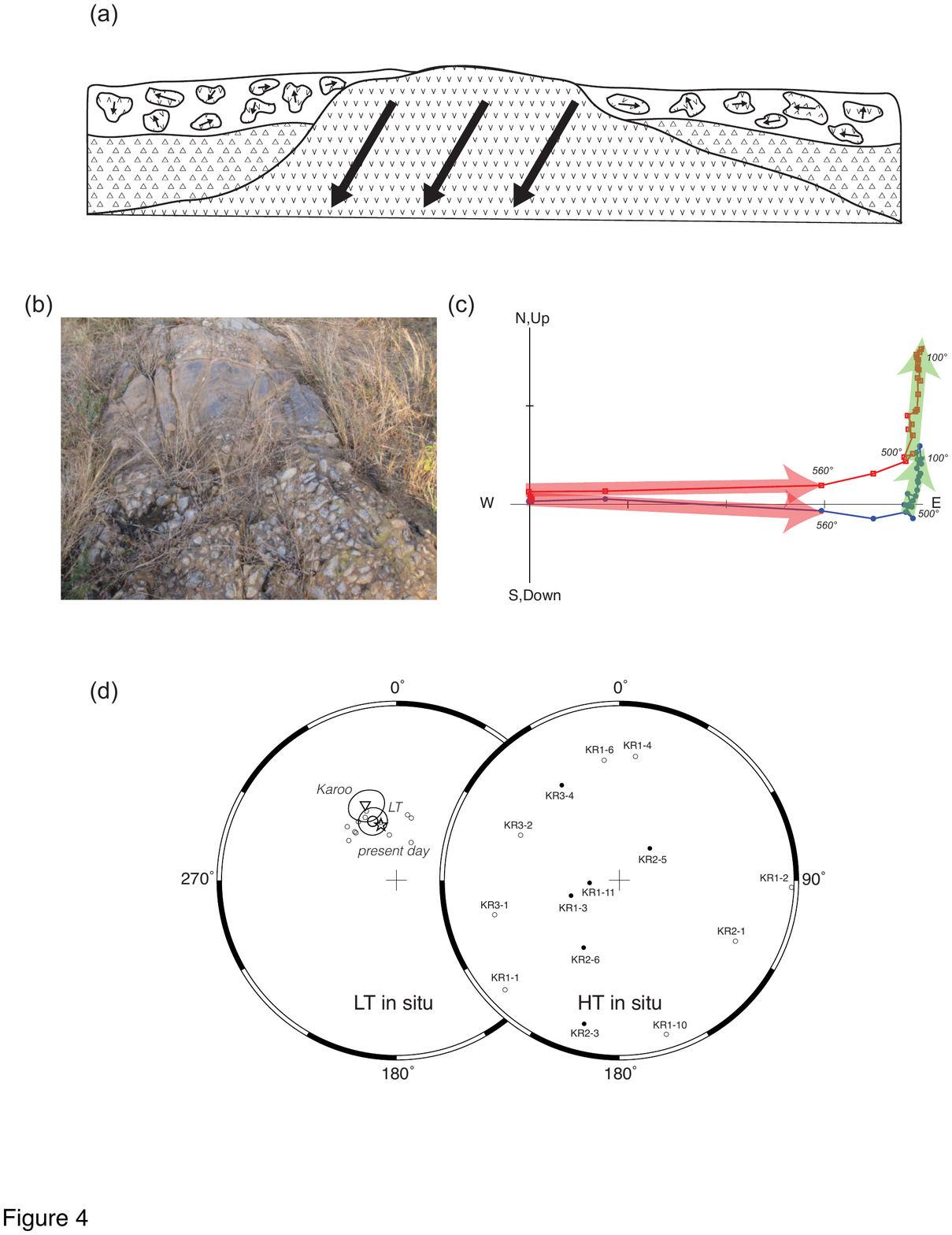}
\caption{
Conglomerate test on $\sim$3.4 Ga rock unit from
the Barberton Greenstone Belt (Kaapvaal Craton). a. Sketch of
conglomerate formation. Key: ``v" symbol, dacite; triangles, chert;
arrows, imparted thermoremanent magnetization. b. Field photo of 3.4
Ga conglomerate. c. Orthogonal vector plot of stepwise thermal
demagnetization of dacite clast from conglomerate; red is inclination,
blue is declination. (b). Two components are defined at low unblocking
temperatures (LT, green) and high unblocking temperatures (HT,
red). d. Stereonet plots show that the LT component is well-grouped,
as expected for a secondary magnetization. The HT component is
scattered, as expected for a primary magnetization (cf. part a). (c)
and (d) are from Usui et al. \citep{Usui2009}.}
\end{figure}

\noindent
\begin{figure}[!ht]
\centering
\includegraphics[width=\linewidth]{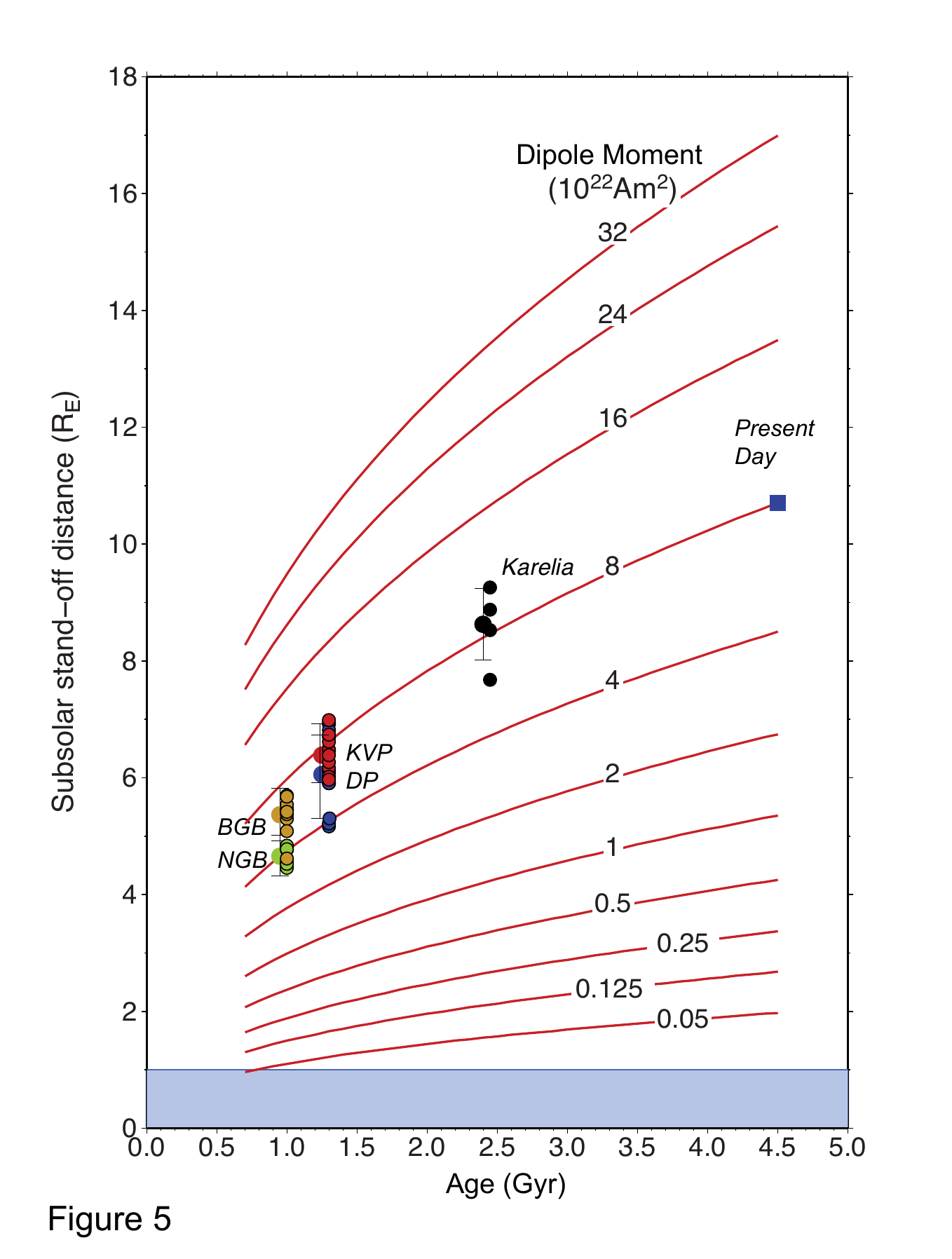}
\caption{Standoff versus time. Subsolar standoff distance in
Earth radii (${\rm R_{E}}$) plotted versus age from solar mass loss
Model A (see text). Contours are Earth's dipole moment, with
paleointensity data from single silicate crystals (cf. Figure 3).
Figure modified from \citep{Tarduno2010}.}
\end{figure}

\noindent
\begin{figure}[!ht]
\centering
\includegraphics[width=\linewidth]{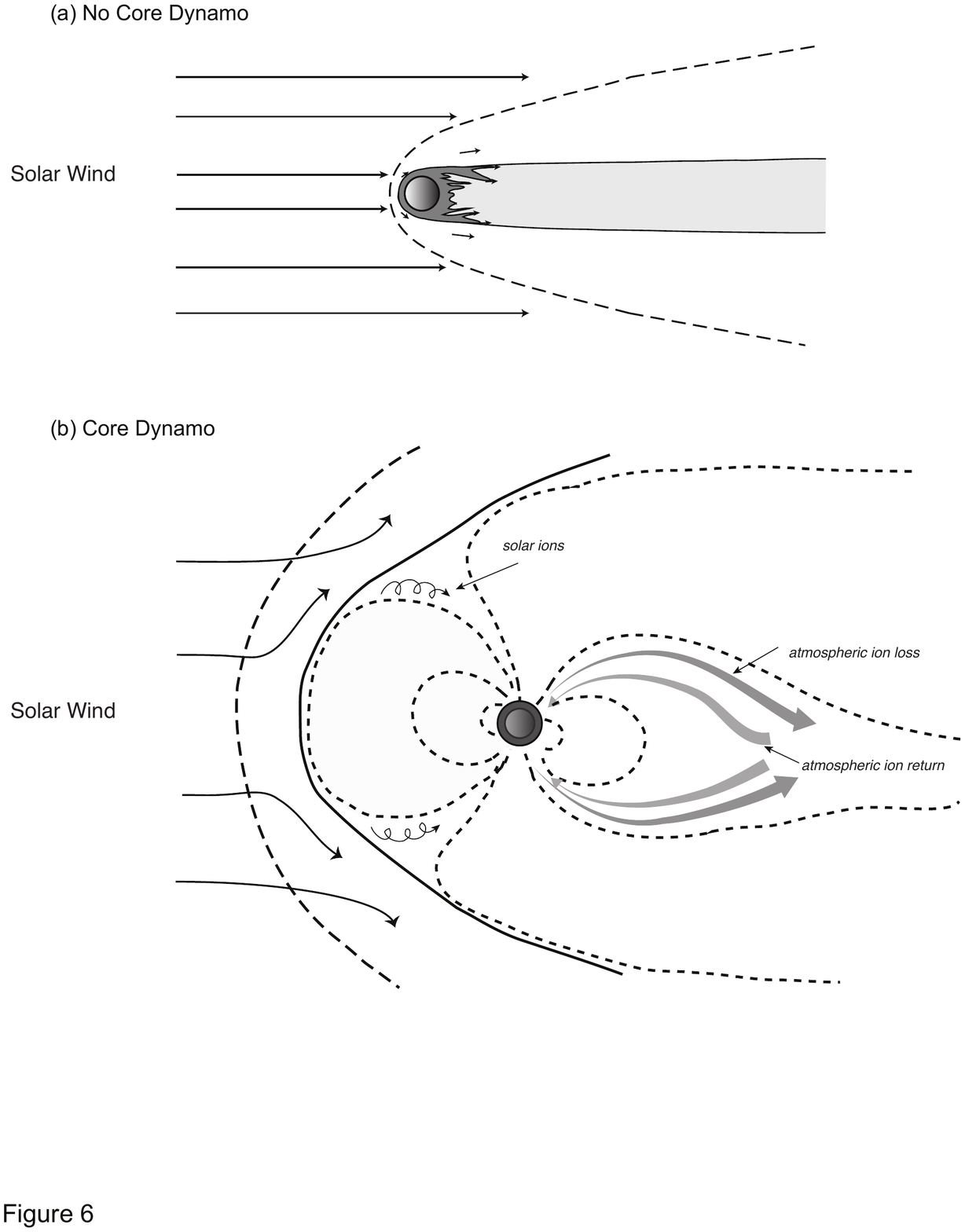}
\caption{Magnetic shielded versus unshielded
planet. a. Planet lacking a core dynamo in the presence of a solar
wind exhibiting atmospheric loss (shaded). b. Planet with a core
dynamo exhibiting atmospheric ion loss and return.}
\end{figure}

\noindent
\begin{figure}[!ht]
\centering
\includegraphics[width=\linewidth]{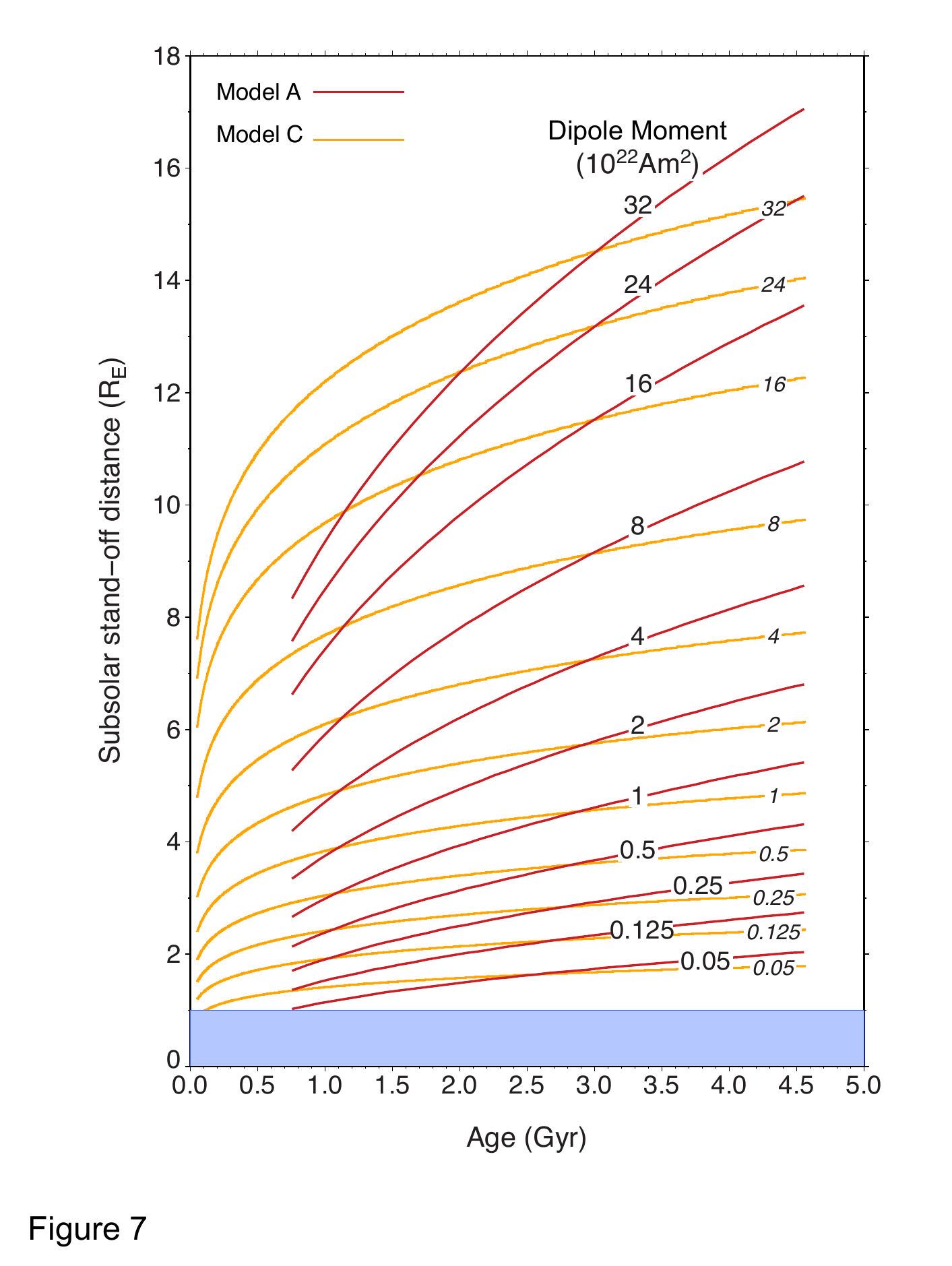}
\caption{Standoff for solar ages younger than 700
Myr. Subsolar standoff distance in Earth radii (${\rm R_{E}}$) plotted
versus age from solar mass loss Model A (red contours) versus Model C
(yellow contours; see text).}
\end{figure}

\noindent
\begin{figure}[!ht]
\centering
\includegraphics[width=\linewidth]{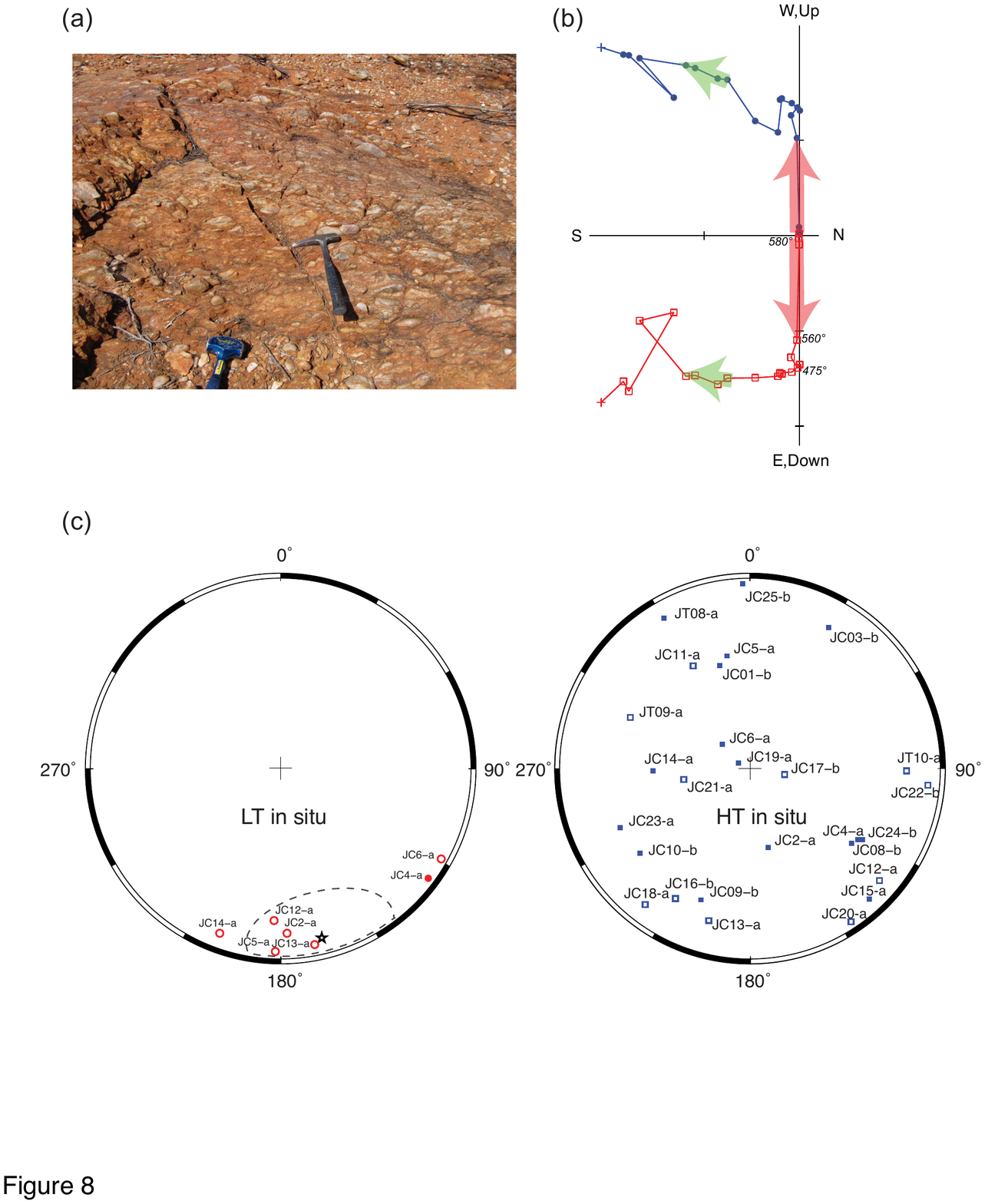}
\caption{Conglomerate test on $\sim$3.0 Ga
  metasediments from the Jack Hills (Yilgarn Craton).  a. Field photo
  of conglomerate with cobble-sized clasts.  b. Orthogonal vector plot
  of stepwise thermal demagnetization of subsample from interior of
  clast from conglomerate; red is inclination, blue is declination.
  Two components are defined at relatively low-intermediate unblocking
  temperatures (LT, green arrows) and high unblocking temperatures
  (HT, red arrows ). Example from \citep{Tarduno2013}. Note that the
  E-W labels on the diagram were inadvertently inverted in the
  original publication.  c. Stereonet plots show that a relatively
  well-grouped LT component from some clasts (left), as expected for a
  secondary magnetization. The HT component is scattered (right), as
  expected for a primary magnetization (from \citep{Usui2009}).}
\end{figure}

\end{document}